\documentclass[10pt, twoside]{IEEEtran}

\usepackage[T1]{fontenc}
\usepackage{amssymb}
\usepackage{amsmath}
\usepackage{setspace}
\usepackage{algorithm}
\usepackage{algorithmicx}

\usepackage{graphicx}
\usepackage{epstopdf}
\DeclareGraphicsExtensions{.eps}

\usepackage{subfigure}
\usepackage{cite}

\makeatletter
 \def\@textbottom{\vskip \z@ \@plus 1pt}
 \let\@texttop\relax
\makeatother

\begin{document}
% can use linebreaks \\ within to get better formatting as desired: Hybrid Precoding and Rate Splitting for Multiuser mmWave Systems with Limited Feedback and Channel Statistics
% Multiuser Transmission for mmWave Systems with Limited Feedback and Channel Statistics
% Hybrid Precoding for Multiuser mmWave Systems with Limited Feedback and Channel Statistics
% Transmission and Precoding Strategy for Multiuser mmWave System with Limited Feedback
\title{Multiuser Millimeter Wave Beamforming Strategies with Quantized and Statistical CSIT}
\author{Mingbo~Dai,~\IEEEmembership{Student Member,~IEEE,}
        Bruno~Clerckx,~\IEEEmembership{Member,~IEEE,}
%\thanks{Manuscript received xxx; revised xxx. The associate editor coordinating the review of this paper and approving it for publication was xxx.

\thanks{%Manuscript received xxx; revised xxx. The associate editor coordinating the review of this paper was xxx.

M. Dai and B. Clerckx are with the Department of Electrical and Electronic Engineering, Imperial College London, SW7 2AZ, UK (e-mail: \{m.dai13, b.clerckx\}@imperial.ac.uk).

This work has been partially supported by the EPSRC of UK, under grant EP/N015312/1.
}}

\maketitle

\begin{abstract}
%\boldmath
To alleviate the high cost of hardware in mmWave systems, hybrid analog/digital precoding is typically employed.
In the conventional two-stage feedback scheme, the analog beamformer is determined by beam search and feedback to maximize the desired signal power of each user.
The digital precoder is designed based on quantization and feedback of effective channel to mitigate multiuser interference.
Alternatively, we propose a one-stage feedback scheme which effectively reduces the complexity of the signalling and feedback procedure.
Specifically, the second-order channel statistics are leveraged to design digital precoder for interference mitigation while all feedback overhead is reserved for precise analog beamforming.
Under a fixed total feedback constraint, we investigate the conditions under which the one-stage feedback scheme outperforms the conventional two-stage counterpart.
Moreover, a rate splitting (RS) transmission strategy is introduced to further tackle the multiuser interference and enhance the rate performance.
Consider (1) RS precoded by the one-stage feedback scheme and (2) conventional transmission strategy precoded by the two-stage scheme with the same first-stage feedback as (1) and also certain amount of extra second-stage feedback.
We show that (1) can achieve a sum rate comparable to that of (2).
Hence, RS enables remarkable saving in the second-stage training and feedback overhead.
% 200 words

\end{abstract}

\begin{IEEEkeywords}
mmWave Systems, Hybrid Precoding, Rate Splitting, Limited Feedback, Statistical CSIT.
\end{IEEEkeywords}

\IEEEpeerreviewmaketitle

\section{INTRODUCTION}
Millimeter wave (mmWave) communication has been recognized as a promising technology in 5G cellular network for its large transmission bandwidth \cite{Rappaport2013, andrews2014}.
To compensate the severe pathloss of mmWave link, large-scale antenna array is needed to provide high precoding gains \cite{Pi2011}.
However, the prohibitive cost and power consumption of radio frequency (RF) chains at mmWave bands makes the fully digital precoding infeasible.
To tackle this RF hardware constraint, a hybrid precoding transceiver architecture has been recently proposed, where the large-scale antenna array is driven by a small number of RF chains \cite{Omar2014}.
The two stage hybrid precoder is implemented by a high-dimensional RF beamformer using cost-efficient analog phase shifters, cascaded with a reduced-dimensional digital precoder \cite{Gaoz2015}.

Considering single user MIMO systems, the hybrid precoder is designed to approach the performance of fully digital precoder by solving a matrix factorization problem \cite{Yu2016}.
When it comes to multiuser MIMO systems, \cite{Foad2016} maximized the sum rate by iteratively optimizing the analog and digital precoder until convergence.
The work \cite{Liang2014} analyzed the rate performance in the large array regime for a given hybrid precoder design.
Furthermore, \cite{Gao2015} considered a partially-connected phase shifter networks.
All these works determined the hybrid precoder assuming perfect full dimensional channel state information at the transmitter (CSIT).

%Nevertheless, the key issue is that during the training phase, what the receiver measures relies on the RF beamforming pattern, which in turn cannot be properly designed without knowing the underlying channel.
%To acquire the channel knowledge, \cite{Alk2014} developed a compressed sensing algorithm while \cite{Bog2015} exploited the knowledge of channel covariance matrix. the impossibility of directly observing the $M$-dimensional channel vector at the antenna array elements due to hybrid precoding architecture. More specifically, \cite{Caire2016} proposed a mmWave channel estimation algorithm that is suitable for the hybrid precoding implementation. To circumvent the coupling effect of RF beamformer design and channel acquisition,
%RVQ is known as the asymptotically optimal codebook in the large array regime when used for beamforming in an independent and identical distributed (i.i.d.) Rayleigh fading channel \cite{Love2004}.
%However, the effective channels at the mmWave bands are unlikely to be i.i.d. due to the highly directional link.
%So far, there has been no investigation on how to enhance the rate performance of multiuser mmWave systems in the presence of the second-order channel statistics.
%In addition, given that the total feedback is limited, how to balance the feedback overhead between two stages is still missing.

In practical mmWave systems, only an imperfect CSIT is attainable through channel estimation \cite{Alk2014, Bog2015} and quantization.
With limited feedback, \cite{Alk2015} proposed a low complexity hybrid precoding approach.
In the first stage, the RF beamformer is designed to maximize the desired signal power of each user by beam search and feedback.
In the second stage, the digital precoder depends on the random vector quantization (RVQ) and feedback of the effective channel (the channel concatenated with the RF beamformer).
This hybrid precoding method relies on two-stage feedback and thus requires a complicated signalling and feedback procedure.
So far, there has been no investigation on how to simplify the signalling and feedback procedure while maintaining the rate performance.

In presence of statistical CSIT, we propose a hybrid precoding design based on one-stage feedback.
Specifically, we make use of all feedback overhead for the first stage to enable precise design of beamforming directions and take advantage of the second-order channel statistics to mitigate multiuser interference.
Hereinafter, this is referred to as One-Stage Feedback plus Statistical CSIT (`OSF + Stat')-based hybrid precoding scheme.
To make a fair comparison, we consider an enhanced design of \cite{Alk2015} by employing a second-order channel statistics-based quantization codebook in the second-stage feedback.
Hereinafter, this is referred to as the Two-Stage Feedback plus Adaptive Codebook (`TSF + Adp CB')-based hybrid precoding scheme.
With a fixed total feedback constraint, we mainly investigate the conditions under which the one-stage feedback scheme outperforms the conventional two-stage counterpart.

Nevertheless, the multiuser interference is still a limiting factor for mmWave systems and leads to a performance degradation.
To alleviate this issue, a rate splitting (RS) transmission strategy was recently proposed in \cite{Hao2013}.
RS relies on superposition coding and successive interference cancellation (SIC) techniques.
Specifically, each user's message is split into a common and a private message\footnote{From sum rate perspective, it does not matter whether RS splits each user's message or a selected subset of users' messages.}, where the common messages are drawn from a public codebook and then packed into one super common message.
This common message can be intended to a subset of the users but should be decoded by all users with zero error probability.
The private messages are transmitted using a fraction of the total power while the common message is superimposed on top of the private messages using the residual power.
At the receiver side, the common message is decoded by treating all the private messages as noise.
After removing the decoded common message from the received signal by SIC, each user decodes their own private messages.
When the CSIT error variance decays with signal-to-noise ratio ($P$) as $O(P^{-\delta})$ for some constant $0 \le \delta < 1$, conventional multiuser transmission strategies with linear precoding (e.g., Zero-Forcing (ZF)) and uniform power allocation achieve a sum Degree-of-Freedom (DoF) of $K\delta$ in $K$-user MISO broadcast channel.
By contrast, the achievable sum DoF of RS is $1 + (K-1)\delta$, which is strictly larger than $K\delta$ achieved with ZF.
In subsequent works, RS has been shown to be a very promising strategy in a wide range of scenarios, namely future PHY layer strategy \cite{Clerckx2016}, massive MIMO \cite{Dai2016}, MIMO networks \cite{Hao2016-1, Hao2016-2}, multi-group multicast \cite{Hamdi2016-3} and under various performance metric, namely sum-rate maximization \cite{Hamdi2016-1}, max-min fairness \cite{Hamdi2016-2}.
However, the benefit of RS in the context of multiuser mmWave systems with hybrid precoding has never been investigated.

Specifically, the main contributions are listed as follows.

\begin{itemize}
  \item With the conventional multiuser transmission strategy, we consider two hybrid precoding schemes each based on one-stage and two-stage feedback, respectively.
  Under a constraint on the total feedback overhead, the `OSF + Stat' scheme exploits the statistical CSIT to design the second-stage digital precoder to mitigate multiuser interference while all feedback resources are reserved to precisely design the first-stage analog beamforming.
  The rate performance of the `OSF + Stat' hybrid precoding scheme is analyzed both in single-path channel and in the large-scale array regime with multiple paths.
  By contrast, the `TSF + Adp CB' scheme allocates partial feedback resources to design the first-stage beamforming while using the residual resources for the effective channel quantization based on an adaptive codebook.
  Then, the second-stage digital precoder is determined based on the quantized CSIT.
  We show that for very limited feedback and/or very sparse channels, the `OSF + Stat' hybrid precoding scheme outperforms the `TSF + Adp CB' counterpart with arbitrary feedback allocation between the two stages.
  In addition, the proposed one-stage feedback scheme effectively reduces the complexity of the signalling and feedback procedure.

  \item In the context of mmWave systems with hybrid precoding, we introduce the RS transmission strategy to further tackle the multiuser interference incurred by imperfect (i.e., quantized/statistical) CSIT.
  Given a certain amount of feedback for the first-stage analog beamforming, the RS strategy with the `OSF + Stat' hybrid precoding can achieve a DoF of 1 (i.e., $\delta = 0$ due to zero second-stage feedback).
  By contrast, the conventional transmission strategy with the `TSF + Adp CB' hybrid precoding requires extra second-stage feedback bits $O(\frac{K-1}{K}\frac{P_{dB}}{3})$ to obtain a DoF of 1.
  It implies that RS can achieve a rate performance comparable to that of conventional No-RS, i.e., only a constant rate gap lies between them.
  Compared with the conventional No-RS strategy, RS enables significant saving in the second-stage training and feedback whereas it requires superposition coding at the transmitter and SIC at the receiver.
  Therefore, we develop two candidate transmission strategies with different requirements for training/feedback and transceiver techniques, which provides design flexibility for different system configurations.

%RS tackles multiuser interference by transmitting an extra common message and can be applied with both aforementioned hybrid precoding schemes.
%RS achieves significant rate gain over conventional No-RS strategies with the same feedback overhead, or equivalently, retains the rate performance with remarkable saving in both channel training and feedback overhead.
%In this paper, we mainly investigate the following problem: how much feedback can be saved by RS compared with the conventional No-RS transmission strategy to achieve comparable rate performance.
%This paper mainly investigates the following problem: In the context of multiuser mmWave systems with hybrid precoding, how much feedback can be saved by RS compared with the conventional No-RS transmission strategy to achieve comparable rate performance.

\end{itemize}

\emph{Organization:} Section \ref{systemmod} introduces the system and the channel models.
Section \ref{BC} elaborates on the proposed hybrid precoding schemes and analyzes the rate performance.
Section \ref{RS} introduces the RS transmission strategy and the associated precoder design and power allocation.
Section \ref{numerical} presents the numerical results and Section \ref{conclusion} concludes the paper.

\emph{Notations:} Bold lower/upper case letters denote vectors and matrices, respectively.
The notations $\mathbf{X}_{i,j}$, $\mathbf{X}^T$, $\mathbf{X}^H$, $\mathbf{X}^{-1}$, $\text{tr}(\mathbf{X})$ denote the entry in the $i$-th row and $j$-th column, the transpose, conjugate transpose, inverse and trace of a matrix $\mathbf{X}$.
The $l_0$-norm and $l_2$-norm of $\mathbf{x}$ are denoted by $\|\mathbf{x}\|_0$ and $\|\mathbf{x}\|$, respectively.
$\mathbf{I}$ is the identity matrix and $\mathbf{1}_i$ is the $i$-th column of $\mathbf{I}$.
The eigenvectors corresponding to the largest and smallest eigenvalues are denoted by $\mathbf{u}_{\text{max}}(\cdot)$ and $\mathbf{u}_{\text{min}}(\cdot)$, respectively.
The operator $\text{diag}(\cdot)$ stands for a diagonal matrix whereas $\mathbb{E}(\cdot)$ represents expectation. We denote $\text{Exp}(c)$ as the exponential distribution with parameter $c$.

\begin{figure*}[t] \centering
\includegraphics[height = 0.32\textwidth, width = 0.8\textwidth]{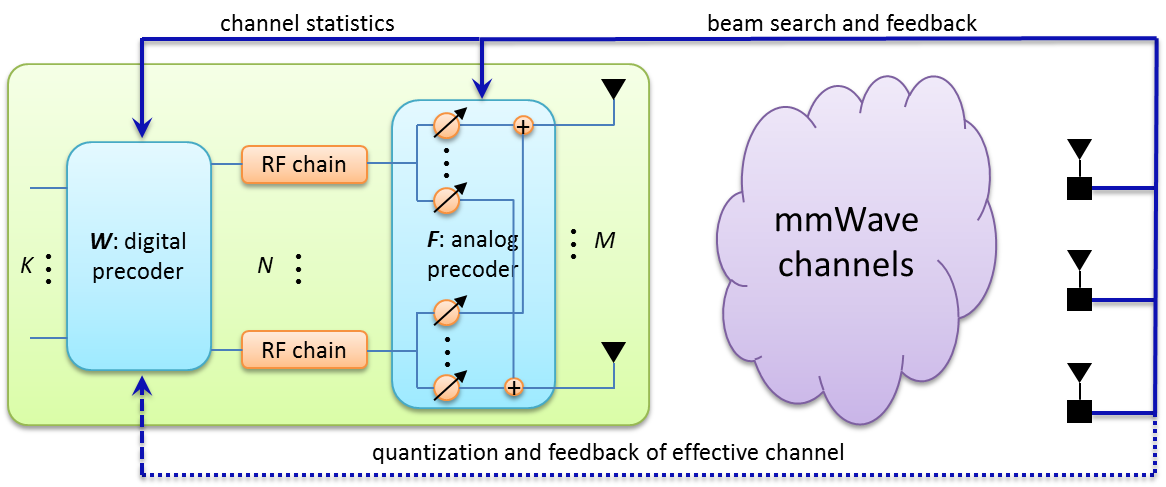}
\caption{Block diagram of multiuser mmWave downlink system model with hybrid precoding and limited feedback. Solid lines indicates the feedback requirement of the hybrid precoding scheme in Section \ref{onestage}. Dash lines represents the additional requirement of the hybrid precoding scheme in Section \ref{twostage}.} \label{diag}
\end{figure*}

\section{SYSTEM MODEL} \label{systemmod}
Consider a multiuser downlink system where the base station (BS) equipped with $M$ antennas and $N$ RF chains serves $K (\le N \le M)$ single-antenna users over mmWave channels.
For simplicity, we assume that the BS only uses $K$ out of $N$ RF chains, which provides a lower bound on the rate performance.
A hybrid RF beamformer $\mathbf{F} \in \mathbb{C}^{M \times K}$ and digital precoder $\mathbf{W} \in \mathbb{C}^{K \times K}$ structure is employed at the BS as depicted in Fig. \ref{diag}.
Since the RF beamformer is implemented using phase shifting networks, a constant modulus constraint is imposed on its entries.
Without loss of generality, we assume that $[\mathbf{F}]_{m,n} = \frac{1}{\sqrt{M}} e^{j\varphi_{m,n}}$.

%%%%%%%%%%%%%%%%%%%%%%%%%%%%%%%%%%%%%%%%%%%%%%%%%%%%%%%% update this in the thesis %%%%%%%%%%%%%%%%%%%%%%%%%%%%%%%%%%%%%%%%%%%%%%%%%%%%%%%%%%
The channels in the mmWave bands tend to be sparse and we assume a ray-based geometric channel model with limited paths \cite{Say2007, Mo2014}.
Under this model, the channel vector from user $k$ is defined as
\begin{eqnarray} \label{eq:channel}
\mathbf{h}_k = \sqrt{\frac{M}{L_k}} \sum^{L_k}_{l=1} g_{k,l} \, \mathbf{a}(\theta_{k,l}) =  \sqrt{\frac{M}{L_k}} \mathbf{A}_k \mathbf{g}_k,
\end{eqnarray}
where the path gain vector $\mathbf{g}_{k} = [g_{k,1}, \cdots, g_{k,L_k}]$ has independent and identical distributed (i.i.d.) $\mathcal{CN} (0,1)$ entries and varies independently across different time slots.
$\mathbf{A}_k = [\mathbf{a}(\theta_{k,1}),\cdots, \mathbf{a}(\theta_{k,{L_k}})] \in \mathbb{C}^{M \times L_k}$ contains $L_k$ steering vectors and $\theta_{k,l} \in [0, \pi]$ are the angle-of-departure (AoD) of $l^{th}$ path.
Accordingly, the long-term channel covariance matrix of $\mathbf{h}_k$ can be computed as
\begin{equation} \label{eq:correlation}
\mathbf{R}_k = \mathbb{E}\left\{ \mathbf{h}_k \mathbf{h}^H_k\right\} = \frac{M}{L_k} \mathbf{A}_k \mathbf{A}^H_k,
\end{equation}
which mainly depends on the AoDs.
It is well known that the fading channel statistics (e.g., AoDs) are wide-sense stationary (WSS) due to its scattering-dependency.
In particular, the AoDs remain invariant over the entire duration of transmission in typical mmWave scenarios \cite{Adh2014}.
Hence, prior to the transmission, the AoD information can be efficiently extracted by channel estimation techniques such as \cite{Alk2014}.
For ease of exposition, $\mathbf{R}_k$ is assumed to be perfectly known at the BS.

%%%%%%%%%%%%%%%%%%%%%%%%%%%%%%%%%%%%%%%%%%%%%%%%%%%%%%%% update this in the thesis %%%%%%%%%%%%%%%%%%%%%%%%%%%%%%%%%%%%%%%%%%%%%%%%%%%%%%%%%%

While the design in this paper is applicable to arbitrary antenna arrays, we consider a uniform linear array (ULA) for ease of exposition.
Under the plane wave and balanced narrowband array assumptions, the array steering vector can be written as
\begin{eqnarray} \label{eq:steer}
\mathbf{a}(\theta_{k,l}) \!=\! \frac{1}{\sqrt{M}} [1, e^{j 2 \pi\frac{d}{\lambda} \cos(\theta_{k,l})},\!\cdots\!,\! e^{j 2 \pi\frac{(M-1)d}{\lambda} \cos(\theta_{k,l})}]^T,
\end{eqnarray}
where $\lambda$ is the wavelength and $d = \frac{\lambda}{2}$ is the antenna spacing.

%The mm-Wave channel estimation has been intensively investigated in \cite{Caire2016, Alk2014}.
%The knowledge of channel state information (CSI) at the receiver (i.e., $\mathbf{h}_k$) is commonly achieved in various wireless standards including mm-Wave standards (e.g., 802.11ad) \cite{Adh2014}.
%Since the AoDs are slowly-varying due to its scattering-dependency, the channel covariance matrix $\mathbf{R}_k$ can be estimated by user $k$ as \cite{Wyne2011}
%\begin{eqnarray} \label{eq:Rkestimation}
%\widehat{\mathbf{R}}_k = \frac{1}{|S|} \sum_{s \in S} \mathbf{h}_{k}(s) \mathbf{h}^H_{k}(s),
%\end{eqnarray}
%where $S$ is the set of channel samples and $|S|$ denotes cardinality of $S$.
%Then, $\widehat{\mathbf{R}}_k$ can be fed back to the BS via a low-rate backhaul link.
%For simplicity, we assume that the BS perfectly knows the long-term channel covariance matrices
%\begin{equation} \label{eq:correlation}
%\mathbf{R}_k = \mathbb{E}\left\{ \mathbf{h}_k \mathbf{h}^H_k\right\} = \frac{M}{L_k} \mathbf{A}_k \mathbf{A}^H_k.
%\end{equation}

\section{MULTIUSER HYBRID PRECODING} \label{BC}
In this section, we consider the conventional multiuser transmission strategy with two hybrid precoding schemes each based on one-stage and two-stage feedback, respectively\footnote{The frameworks in this work are primarily envisioned for Frequency Division Duplex (FDD) systems.}.
Under a fixed total feedback overhead constraint\footnote{Since both of the hybrid precoding schemes under consideration utilize the second-order channel statistics $\mathbf{R}_k$, the feedback overhead of $\mathbf{R}_k$ can be neglected in terms of the comparison between them.},
we explore the conditions under which the `OSF + Stat' hybrid precoding scheme outperforms the `TSF + Adp CB' counterpart.
Moreover, the rate performance of the `OSF + Stat' scheme is analyzed both in single-path channel and in the large-scale array regime with multiple paths.

In the conventional multiuser transmission strategy, the transmitted signal and the received signal of user $k$ can be written as
\begin{eqnarray} \label{eq:tx_sigbc}
\mathbf{x} &=& \mathbf{F} \mathbf{W} \mathbf{P} \mathbf{s} = \sum^K_{k = 1}  \sqrt{P_{k}} \mathbf{F} \mathbf{w}_{k} \, s_{k}, \\ \label{eq:rx_sigbc}
y_k &=& \sqrt{P_{k}} \, \mathbf{h}^H_{k} \mathbf{F} \mathbf{w}_{k} \, s_{k} + \underbrace{\sqrt{P_{j}} \mathbf{h}^H_{k} \sum_{j \ne k}^{K} \mathbf{F} \mathbf{w}_{j} \, s_{j}}_{\text{multiuser interference}} + n_k,
\end{eqnarray}
where $\mathbf{s} = [s_1, \cdots, s_K]^T \in \mathbb{C}^K$ is the data vector intended for the $K$ users. The transmit power is uniformly allocated among users, i.e., $\mathbf{P} = \sqrt{P/K} \cdot \mathbf{I}_{K}$ and $\|\mathbf{F} \mathbf{w}_{k}\|^2 = 1$. $n_k \sim \mathcal{CN} (0,1)$ is the additive white Gaussian noise (AWGN). Then, the signal-to-interference-and-noise-ratio (SINR) of user $k$ is computed as
\begin{eqnarray} \label{eq:sinr1}
\text{SINR}_{k} =  \frac{\rho \, |\mathbf{h}^H_{k,\text{eff}} \mathbf{w}_{k} |^2 }{1+ \rho \sum_{j \ne k} |\mathbf{h}^H_{k,\text{eff}} \mathbf{w}_{j} |^2 },
%\frac{\frac{P}{K} \, |\mathbf{h}^H_{k} \mathbf{F} \mathbf{w}_{k} |^2 }{1+ \frac{P}{K} \sum_{j \ne k} |\mathbf{h}^H_{k} \mathbf{F} \mathbf{w}_{j} |^2 } =
\end{eqnarray}
where we define the effective channel by $\mathbf{h}_{k,\text{eff}} = \mathbf{F}^H \mathbf{h}_{k} \in \mathbb{C}^{K \times 1}$ and $\rho = P/K$. The sum rate can be written as $R_{\text{sum}} = \sum^{K}_{k=1} \log_2 (1 + \text{SINR}_{k})$. To facilitate a low complexity design, $\mathbf{F}$ and $\mathbf{W}$ are determined in a decoupled manner \cite{Alk2015}.

\subsection{One-Stage Feedback Scheme} \label{onestage}
The `OSF + Stat' hybrid precoding scheme uses all feedback resources to precisely design the first-stage analog beamsteering while exploiting the statistical CSIT to mitigate multiuser interference.
Let $\mathcal{F}$ represent the RF beamsteering codebook, where $\mathcal{F} = \big\{ \mathbf{a}(\theta_q)| \theta_q = \frac{\pi q}{Q}, q \in [1, Q] \big\}$ with cardinality $|\mathcal{F}| = Q = 2^B$.
For each channel realization, the BS searches beams in the codebook and user $k$ feeds back the index of the codeword that gives the maximum received power.
Efficient beam search algorithm can be found in \cite{Xiao2016} and references therein.
Then, the BS sets the selected codeword as $\mathbf{f}_k$, i.e.,
\begin{eqnarray} \label{eq:beam}
\{\mathbf{f}_k\} = \underset{\mathbf{f}_k \in \mathcal{F} }{\text{arg max}} \;\; |\mathbf{h}^H_k \mathbf{f}_k|^2.
\end{eqnarray}

We assume full rank $\mathbf{F}$ (i.e., users have different dominant paths and the BS has distinct beamforming direction for each user).
The probability of this event approaches one for independently and randomly distributed AoDs, large AoDs space and feedback overhead $B = \log_2(M)$\footnote{The total feedback overhead $B = 6 \sim 8$ can support a large antenna array with dimension $64 \sim 256$ and therefore this assumption is valid in practice.}.

As $\mathbf{F}$ and $\mathbf{R}_k$ are known to the BS, the BS can equivalently compute the covariance matrix of the effective channel $\mathbf{h}_{k,\text{eff}} = \mathbf{F}^H \mathbf{h}_{k} \in \mathbb{C}^{K \times 1}$ as $\mathbf{R}_{k,\text{eff}} = \mathbf{F}^H \mathbf{R}_{k} \mathbf{F} \in \mathbb{C}^{K \times K}$.
$\mathbf{R}_{k,\text{eff}}$ can be utilized to mitigate the multiuser interference.
Noting that $\mathbf{W} \in \mathbb{C}^{K \times K}$, a straightforward design of the digital precoder $\mathbf{w}_k$ of user $k$ lies in the nullspace of $\mathcal{S}_k$, denoted as statistical beamforming (SBF), i.e.,
\begin{eqnarray} \label{eq:dgprecoder}
\mathbf{w}_k = \text{Null} \{ \mathcal{S}_k  \},
\end{eqnarray}
where $\mathcal{S}_k = \text{Span}(\{\mathbf{u}_{\text{max}}(\mathbf{R}_{j,\text{eff}}): j \ne k \})$ is defined as the space spanned by the dominant eigenvectors of $K \times K$ channel covariance matrices $\mathbf{R}_{j,\text{eff}}$ of all other users $j \ne k$.
This design intends to minimize the multiuser interference in a statistical sense but overlooks the desired signal power.
To overcome this problem, we adopt a signal-to-leakage-and-noise ratio (SLNR) metric which strikes a balance between the desired signal power and the interference imposed to other users
\begin{eqnarray} \label{eq:slnr}
\text{SLNR}_{k} = \frac{\rho \, |\mathbf{h}^H_{k,\text{eff}} \mathbf{w}_{k} |^2 }{1+ \rho \sum_{j \ne k} |\mathbf{h}^H_{j,\text{eff}} \mathbf{w}_{k} |^2 }.
\end{eqnarray}

The SLNR metric has been widely used for designing multiuser transmit beamforming \cite{Sadek2007,Raghavan2011}.
Then, $\mathbf{W}$ is designed by maximizing a lower bound on the average SLNR
\begin{eqnarray} \label{eq:slnr_lb}
\mathbb{E}(\text{SLNR}_{k}) &=& \mathbb{E}(|\mathbf{h}^H_{k,\text{eff}} \mathbf{w}_{k} |^2 ) \mathbb{E}\big(\frac{1}{\frac{1}{\rho} + \sum_{j \ne k} |\mathbf{h}^H_{j,\text{eff}} \mathbf{w}_{k} |^2 } \big) \nonumber \\ \label{eq:slnr_lb2}
&\ge& \frac{\mathbf{w}^H_{k} \mathbf{R}_{k,\text{eff}} \mathbf{w}_{k} }{\frac{1}{\rho}+\sum_{j \ne k} \mathbf{w}^H_{k} \mathbf{R}_{j,\text{eff}} \mathbf{w}_{k} } \triangleq \text{SLNR}^{\text{LB}}_{k},
\end{eqnarray}
where the first equation leverages independence between the numerator and denominator of SLNR and \eqref{eq:slnr_lb2} is obtained by the convexity of $f(x) = 1/x$.
The optimal unit norm $\mathbf{w}_k$ that maximizes the lower bound in \eqref{eq:slnr_lb} is the generalized eigenvector given by \cite{Dai2015}
\begin{eqnarray} \label{eq:GE}
\mathbf{w}_{k} = \mathbf{u}_{\text{max}} \Big( \big( \frac{1}{\rho} \mathbf{I} + \sum_{j \ne k} \mathbf{R}_{j,\text{eff}} \big)^{-1} \mathbf{R}_{k,\text{eff}} \Big).
\end{eqnarray}

Subject to the total transmit power constraint, the digital precoder is then normalized as $\mathbf{w}_{k} = \frac{\mathbf{w}_{k}}{\|\mathbf{F} \mathbf{w}_{k}\|}$.
The `OSF + Stat' hybrid precoding scheme is summarized in Algorithm \ref{hyalg1}.
Next, we analyze the achievable sum rate in single-path channels and in the large array regime respectively.
The analysis in these special cases gives insights into the rate performance of more general settings.

\begin{algorithm}[t]
\caption{: Hybrid Precoding Schemes}\label{hyalg1}
\begin{algorithmic}[1]
\State \textbf{Input}: `OSF': RF codebook $\mathcal{F}$ of size $|\mathcal{F}|=2^B$ or `TSF': RF codebook $\mathcal{F}$ of size $2^{B_{RF}}$ and statistical CSIT-based digital codebook $\mathcal{C}_k$ of size $|\mathcal{C}_k|=2^{B_{BB}}$
\State \textbf{First stage}: Single-user RF beamforming ($\mathbf{F}$)
    \State \quad Downlink beam search and uplink feedback with \eqref{eq:beam}
\State \textbf{Second stage}: Multiuser digital precoding ($\mathbf{W}$)
    \State \quad `OSF': SBF \eqref{eq:dgprecoder} or SLNR-based \eqref{eq:GE} or
    \State \quad `TSF': channel quantization and feedback with \eqref{eq:quant} and
                 ZF \eqref{eq:zfbf} or SLNR-based \eqref{eq:GE2}
\end{algorithmic}
\end{algorithm}

\subsubsection{Single-path channels}
When $L_k = 1, \forall k$, the channel covariance matrix $\mathbf{R}_k = M \mathbf{a}_k \mathbf{a}^H_k$ is rank one.
Equivalently, the direction of the instantaneous channel is perfectly known at the BS.
The composite channel matrix can be written as $\mathbf{H} = \mathbf{A} \cdot M \text{diag} \{g_1, \cdots, g_K \}$ with $\mathbf{A} \triangleq [\mathbf{a}_1, \cdots, \mathbf{a}_K]$.
To maximize the desired signal power, the BS steers beams to each user via matched beamforming, i.e., $\mathbf{F} = \mathbf{A}$.
Then, the effective channel matrix becomes $\bar{\mathbf{H}}_{\text{eff}} = \mathbf{A}^H \mathbf{A}$ with each column $\bar{\mathbf{h}}_{k,\text{eff}} = \mathbf{A}^H \mathbf{a}_k$. The digital precoder is designed as ZF, i.e., $\mathbf{W} = \bar{\mathbf{H}}^{-1}_{\text{eff}} \mathbf{\Sigma}  =  (\mathbf{\mathbf{A}^H \mathbf{A}})^{-1} \mathbf{\Sigma}$, where the diagonal matrix $\mathbf{\Sigma}$ contains the normalization factor to make $\|\mathbf{F} \mathbf{w}_{k}\| = 1, \forall k$. Thus, it is easily to verify that $\mathbf{\Sigma}_{k,k} = \frac{1}{(\mathbf{\mathbf{A}^H \mathbf{A}})^{-1}_{k,k}}$. The achievable rate of user $k$ is given by
\begin{eqnarray} \label{eq:rate1}
R_k &=& \log_2 \Big(1 + \frac{\rho M |g_k|^2 \cdot |\mathbf{a}^H_{k} \mathbf{F} \mathbf{w}_{k} |^2 }{1+ \rho M |g_k|^2 \cdot \sum_{j \ne k} |\mathbf{a}^H_{k} \mathbf{F} \mathbf{w}_{j} |^2}\Big) \nonumber \\
 &=& \log_2 \Big(1 + \frac{\rho M |g_k|^2 \cdot |\bar{\mathbf{h}}^H_{k,\text{eff}} \mathbf{w}_{k} |^2 }{1+ \rho M |g_k|^2 \cdot \sum_{j \ne k} |\bar{\mathbf{h}}^H_{k,\text{eff}} \mathbf{w}_{j} |^2}\Big) \nonumber \\
&=& \log_2 \Big(1 + \frac{\rho M |g_k|^2 }{(\mathbf{\mathbf{A}^H \mathbf{A}})^{-1}_{k,k} }\Big),
\end{eqnarray}
which coincides with \cite[Theorem 1]{Alk2015}. \cite{Alk2015} assumes continuous angles of the RF beamsteering vectors and perfect effective channel knowledge, i.e., both the RF codebooks and the effective channel quantization codebooks are assumed with infinite resolution.
By contrast, we assume perfect statistical CSIT.
It implies that in single-path mmWave channels, second-order channel statistics are sufficient to achieve the same rate as that achievable by infinite resolution codebooks.

\subsubsection{Large array regime}
For tractability, the rate analysis of multi-path channel is based on the assumption of a ULA with a large number of transmit antennas. In this case, the channel model \eqref{eq:channel} can be well approximated by virtual channel representation (VCR) \cite[Ch.~3]{Bruno2013}
\begin{eqnarray} \label{eq:channel2}
\mathbf{h}_k = \sqrt{\frac{M}{L}} \sum^{L}_{l=1} g_{k,l} \, \mathbf{e}_{k,l} = \sqrt{\frac{M}{L}} \mathbf{E}_k \mathbf{g}_k,
\end{eqnarray}
where $L_k = L, \forall k$ is assumed for simplicity. The steering vectors $\mathbf{E}_k$ contains $L$ columns of the DFT matrix $\mathbf{E}$. Without loss of generality, we assume the path gains in descending order $|g_{k,1}| \ge ... \ge |g_{k,L}|$. The corresponding channel covariance matrix is expressed as
\begin{eqnarray} \label{eq:cov}
\mathbf{R}_k = \frac{M}{L} \mathbf{E}_k \mathbf{E}^H_k = \frac{M}{L} \mathbf{E} \mathbf{D}_k \mathbf{E}^H,
\end{eqnarray}
where $ \mathbf{D}_k = \text{diag}\{\mathbf{d}_k\}$, $\mathbf{d}_k \in \{0,1\}$ and $\|\mathbf{d}_k\|_0 = L$ (i.e., rank($\mathbf{R}_k$) $= L$).
Below, we analyze the achievable rate of the proposed `OSF + Stat' scheme in two special cases.

\emph{Non-overlapped}: the AoD spread and user locations are well separated such that users exhibit non-overlapped multi-paths. %mutually

\emph{Fully overlapped}: users are confined into an area such that all users share the same AoDs and thus the same channel covariance matrix $\mathbf{R}$.
%Since the number of interference-free streams that the BS can emit (i.e., the number of single-antenna users that the BS can serve simultaneously) is limited by the rank of $\mathbf{R}$ (i.e., $L$), we assume the number of paths no less than the number of users, i.e., $L \ge K$.

$\textbf{Proposition 1:}$ Based on \eqref{eq:GE}, the achievable rate of user $k$ of the proposed `OSF + Stat' hybrid precoding scheme in the large array regime is respectively given as
\begin{eqnarray} \label{eq:achrate1}
R_k = \log_2 \Big(1 + \rho\frac{M}{L} |g_{k,1}|^2 \Big)
\end{eqnarray}
in the non-overlapped scenario, and as
\begin{eqnarray} \label{eq:achrate2}
%&=& \log_2 \Big(1 + \frac{\rho\frac{M}{L} |g_{k,1}|^2 }{1 + \rho\frac{M}{L} \sum_{j \ne k} |g_{k,l_j(l_j \ne 1)}|^2} \Big) \\
R_k  \ge \log_2 \Big(1 + \frac{\rho\frac{M}{L} |g_{k,1}|^2 }{1 + \rho\frac{M}{L} \sum^L_{l=2} |g_{k,l}|^2} \Big)
\end{eqnarray}
in the fully overlapped scenario.
\begin{IEEEproof}
See Appendix \ref{sec:prop1}.
\end{IEEEproof}

\emph{Remark 1}: When the AoDs of each user are non-overlapped, \eqref{eq:achrate1} shows that statistical CSIT is able to completely remove the multiuser interference.
\eqref{eq:achrate1} also serves as the interference-free per-user rate $R^s_k$ with $\rho = P/K$.
When the channels are very sparse (i.e., small $L$) or have dominant path (i.e., $|g_1| \gg |g_l|, \forall l \ge 2$), we observe in \eqref{eq:achrate2} that the multiuser interference term $\sum^L_{l=2} |g_{k,l}|^2$ can be neglected at practical SNR.
It implies that the long-term channel statistics enables efficient interference nulling in above scenarios.
By contrast, the closed-form achievable rate in partially overlapped AoDs scenario cannot be easily attained due to the complicated structure of $\mathbf{w}_k$ in \eqref{eq:GE}.
Nevertheless, we note that $R_k$  is upper and lower bounded by \eqref{eq:achrate1} and \eqref{eq:achrate2}, respectively.
To get insights into the rate performance of `OSF + Stat' in partially overlapped AoDs scenario, we quantify the rate gap between \eqref{eq:achrate1} and \eqref{eq:achrate2} as follows.
%From Appendix \ref{sec:prop1}, the hybrid precoding in the special non-overlapped/fully-overlapped cases boils down to the RF-only beamsteering (i.e., $\mathbf{W} = \mathbf{I}$).

$\textbf{Corollary 1:}$ For the proposed `OSF + Stat' hybrid precoding scheme in the large array regime, the average rate gap $\Delta R_k$ between the worst-case (fully overlapped AoDs) and best-case (non-overlapped AoDs) is upper bounded by
\begin{eqnarray} \label{eq:rategap}
\Delta R_k  \le \log_2 \Big(1 +  \rho M \frac{L-1}{L} \Big).
\end{eqnarray}
\begin{IEEEproof}
According to \eqref{eq:achrate1} and \eqref{eq:achrate2}, we have
\begin{eqnarray} \label{eq:rategapproof}
\Delta R_k  \!\!\!&\le&\!\!\!\! \mathbb{E} \Big[\log_2 \big(1 \!+\! \rho\frac{M}{L} |g_{k,1}|^2 \big) + \log_2 \Big(1 \!+\! \rho\frac{M}{L}\! \sum^L_{l=2} |g_{k,l}|^2 \Big)\nonumber \\
&-&\!\!\! \log_2 \Big(1 + \rho\frac{M}{L} \sum_{l} |g_{k,l}|^2 \Big) \Big]   \\
&\le& \mathbb{E} \Big[\log_2 \Big(1 \!+\! \rho\frac{M}{L}\! \sum^L_{l=2} |g_{k,l}|^2 \Big)\Big].
\end{eqnarray}
By further using Jensen's inequality, \eqref{eq:rategap} is obtained.
\end{IEEEproof}

Corollary 1 shows the upper bound on the rate gap between arbitrary overlapped scenarios and interference-free single-user transmission.
This rate gap upper bound increases with $\rho$, which suggests that the statistical CSIT can barely mitigate multiuser interference in the fully overlapped AoDs case.
Nevertheless, the AoDs of users in mmWave systems are sparse and randomly distributed.
Thus, AoDs are unlikely to be heavily overlapped.
The proposed `OSF + Stat' hybrid precoding scheme can still effectively mitigate the multiuser interference, as shown in Section \ref{Numerical_MU}.

\subsection{Two-Stage Feedback Scheme} \label{twostage}
Under a fixed total feedback overhead constraint, the `TSF + Adp CB' hybrid precoding scheme partitions the feedback into two stages.
The second-stage feedback ($B_{BB}$) is used to effectively quantize the channel $\mathbf{h}_{k,\text{eff}}$.
Rather than using RVQ\footnote{RVQ is known as the asymptotically optimal codebook in the large array regime when used for beamforming in an i.i.d. Rayleigh fading channel \cite{Love2004}.
However, the effective channels at the mmWave bands are unlikely to be i.i.d. due to the highly directional link.} codebook as in \cite{Alk2015}, we employ a statistical CSIT adaptive codebook which has been intensively discussed in the literature \cite{Bruno2008, Choi2013}.
Then, the digital precoder is computed based on the effective channel quantization and thus more capable to mitigate multiuser interference than statistical CSIT-based digital precoder in `OSF + Stat' scheme.
However, the residual feedback ($B_{RF} = B - B_{BB}$) can only enable coarse first-stage analog beamforming and may lead to an undesired rate performance.

%By downlink training and estimation, $\mathbf{h}_{k,\text{eff}}$ is assumed perfectly known at user $k$.
%Moreover, the channel quality information (CQI) (i.e., the channel gain $\|\mathbf{h}_{k,\text{eff}}\|$) is perfectly known to BS while
The RF beamforming follows \eqref{eq:beam} with $|\mathcal{F}| = 2^{B_{RF}}$ and the effective channel is given as $\mathbf{h}_{k,\text{eff}} = \mathbf{F}^H \mathbf{h}_{k}$.
The channel quality information (CQI) (i.e., the effective channel gain $\|\mathbf{h}_{k,\text{eff}}\|$) is assumed perfectly known to the BS while the channel direction information (CDI) $\bar{\mathbf{h}}_{k,\text{eff}}= \mathbf{h}_{k,\text{eff}}/\|\mathbf{h}_{k,\text{eff}}\|$ is quantized, denoted as $\hat{\bar{\mathbf{h}}}_{k,\text{eff}}$, and then fed back to the BS.
A classical channel matching metric is adopted for the effective channel quantization, where each user selects the best codeword that maximizes the inner-product with its channel direction, i.e.,
\begin{eqnarray} \label{eq:quant}
\widehat{\bar{\mathbf{h}}}_{k,\text{eff}} = \underset{\mathbf{c}_k \in \mathcal{C}_k}{\text{arg max}} \;\; |\bar{\mathbf{h}}^H_{k,\text{eff}} \, \mathbf{c}_k|^2.
\end{eqnarray}

According to \eqref{eq:correlation}, the geometric multi-path channel model \eqref{eq:channel} can be written as spatially correlated model $\mathbf{h}_k = \mathbf{R}^{1/2}_k \mathbf{g}_k$.
Then, the effective channel can be equivalently treated as $\mathbf{h}_{k,\text{eff}} = \mathbf{R}^{1/2}_{k,\text{eff}} \;\mathbf{g}_k$ with positive semi-definite (PSD) $\mathbf{R}_{k,\text{eff}} = \mathbf{F}^H \mathbf{R}_{k} \mathbf{F}$.
In this setting, the channel space is no longer a hypersphere in $\mathbb{C}^K$ but a hyperellipse stretched by the eigenvalues of $\mathbf{R}_{k,\text{eff}}$.
Hence, a skewed codebook is preferable which multiplies $\mathcal{C}_{\text{iid}} = \{\mathbf{c}_1, \cdots, \mathbf{c}_{2^{B_{BB}}} \}$ (that is specified to i.i.d. channels) to $\mathbf{R}_{k,\text{eff}}^{1/2}$ \cite{Love2006}, i.e.,
\begin{eqnarray} \label{eq:skewed}
\mathcal{C}_k = \Bigg\{\frac{\mathbf{R}^{1/2}_{k,\text{eff}}\, \mathbf{c}_{i}}{\| \mathbf{R}^{1/2}_{k,\text{eff}}\, \mathbf{c}_{i} \|} \Bigg\}.
\end{eqnarray}

%$\hat{\mathbf{h}}_{k,\text{eff}} = \|\mathbf{h}_{k,\text{eff}}\| \cdot \hat{\bar{\mathbf{h}}}_{k,\text{eff}} $
With the knowledge of $\widehat{\bar{\mathbf{h}}}_{k,\text{eff}} $, $\mathbf{W}$ can be designed as the ZF precoding, i.e.,
\begin{eqnarray} \label{eq:zfbf}
\mathbf{W} = \widehat{\bar{\mathbf{H}}}_{\text{eff}} \, ( \widehat{\bar{\mathbf{H}}}_{\text{eff}}^H \, \widehat{\bar{\mathbf{H}}}_{\text{eff}} )^{-1},
\end{eqnarray}
where $\widehat{\bar{\mathbf{H}}}_{\text{eff}} =[\widehat{\bar{\mathbf{h}}}_{1,\text{eff}} , \cdots, \widehat{\bar{\mathbf{h}}}_{K,\text{eff}} ]$.
Moreover, inspired by the statistical CSIT SLNR-based design of $\mathbf{W}$ in Section \ref{onestage}, we here derive an imperfect CSIT SLNR-based design for a fair comparison between the proposed two schemes.
Briefly, the SLNR based on the quantized CSIT $\widehat{\mathbf{h}}_{k,\text{eff}} = \|\mathbf{h}_{k,\text{eff}}\| \cdot \widehat{\bar{\mathbf{h}}}_{k,\text{eff}}$ writes as
\begin{eqnarray} \label{eq:slnrimp}
\text{SLNR}_{k} &=& \frac{\rho  \,|\widehat{\mathbf{h}}^H_{k,\text{eff}} \mathbf{w}_{k} |^2 }{1+ \rho \sum_{j \ne k}|\widehat{\mathbf{h}}^H_{j,\text{eff}} \mathbf{w}_{k} |^2 } \nonumber \\
&=& \frac{\mathbf{w}^H_{k} \widehat{\mathbf{h}}_{k,\text{eff}} \widehat{\mathbf{h}}^H_{k,\text{eff}} \mathbf{w}_{k}  }{\mathbf{w}^H_{k} \big(\frac{1}{\rho} \mathbf{I}+ \sum_{j \ne k} \widehat{\mathbf{h}}_{j,\text{eff}} \widehat{\mathbf{h}}^H_{j,\text{eff}} \big) \mathbf{w}_{k} },
\end{eqnarray}
and the optimal unit norm $\mathbf{w}_k$ that maximizes \eqref{eq:slnrimp} is the generalized eigenvector computed as \cite{Dai2015}
\begin{eqnarray} \label{eq:GE2}
\mathbf{w}_{k} &=& \mathbf{u}_{\text{max}} \Big( \big( \frac{1}{\rho} \mathbf{I} + \sum_{j \ne k}  \widehat{\mathbf{h}}_{j,\text{eff}} \widehat{\mathbf{h}}^H_{j,\text{eff}} \big)^{-1} \widehat{\mathbf{h}}_{k,\text{eff}} \widehat{\mathbf{h}}^H_{k,\text{eff}} \Big) \nonumber \\
&=& \Big( \frac{1}{\rho} \mathbf{I} + \sum_{j \ne k}  \widehat{\mathbf{h}}_{j,\text{eff}} \widehat{\mathbf{h}}^H_{j,\text{eff}} \Big)^{-1} \widehat{\mathbf{h}}_{k,\text{eff}}.
\end{eqnarray}

Equation \eqref{eq:GE2} is obtained as follows. We first denote $\mathbf{S} = \Big( \frac{1}{\rho} \mathbf{I} + \sum_{j \ne k}  \widehat{\mathbf{h}}_{j,\text{eff}} \widehat{\mathbf{h}}^H_{j,\text{eff}} \Big)^{-1}$ and $\mathbf{x} = \widehat{\mathbf{h}}_{k,\text{eff}}$. Following the definition of eigenvalue $\lambda$ and eigenvector $\mathbf{v}$ of a square matrix $\mathbf{M}$, i.e., $\mathbf{M} \mathbf{v} = \lambda \mathbf{v}$, we can easily find that $\mathbf{M} = \mathbf{S} \mathbf{x} \mathbf{x}^H$, $\mathbf{v} = \mathbf{S} \mathbf{x}$ and $\lambda = \mathbf{x}^H \mathbf{S} \mathbf{x}$. Consequently, $\mathbf{S} \mathbf{x}$ is the eigenvector of $\mathbf{S} \mathbf{x} \mathbf{x}^H$. Since $\mathbf{S} \mathbf{x} \mathbf{x}^H$ has rank of one, $\mathbf{S} \mathbf{x}$ is the largest (non-zero) eigenvector.

Then, we normalize it as $\mathbf{w}_{k} = \frac{\mathbf{w}_{k}}{\|\mathbf{F} \mathbf{w}_{k}\|}$.
The `TSF + Adp CB' hybrid precoding scheme is summarized in Algorithm \ref{hyalg1}.
Next, we analyze the effect of $B_{BB}$ on the sum DoF (also referred to as multiplexing gain).
Based on the design of \eqref{eq:quant} $\sim$ \eqref{eq:zfbf}, when $B_{BB}$ is small and independent of SNR, the achievable rate of the system eventually saturates (i.e., DoF = 0).
In order to achieve a non-zero DoF, $B_{BB}$ should scale at a rate linearly increasing with SNR as follows.

$\textbf{Proposition 2:}$ If $B_{BB}$ is scaled as $B_{BB} = O(\alpha \log_2 P)$ for $\alpha \le r - 1$, the `TSF + Adp CB' hybrid precoding scheme achieves at least a sum DoF of $K \cdot \frac{\alpha}{r-1}$.

\begin{IEEEproof}
As the analog beamformer is determined by \eqref{eq:beam}, the effective channel can be equivalently treated as spatially correlated channel $\mathbf{h}_{k,\text{eff}} = \mathbf{R}^{1/2}_{k,\text{eff}} \;\mathbf{g}_k$. Then, the `TSF + Adp CB' hybrid precoding scheme boils down to the conventional digital precoding scheme.
According to \cite[Lemma 1]{Shen2016},\cite{Bruno2008}, the expected quantization error of the effective channel can be upper bounded by $2^{-B_{BB}/(r-1)}$, where $r$ is the rank of the $K \times K$ covariance matrix of the effective channel $\mathbf{R}_{k,\text{eff}}$ and $r \le K$.
By straightforwardly following the proof of \cite[Theorem 4]{Jindal2006}, we can lower bound the DoF of each user by $\frac{\alpha}{r-1}$ for $B_{BB} = O(\alpha \log_2 P)$.
\end{IEEEproof}
Intuitively, in the asymptotical SNR regime $P \rightarrow \infty$, the signal power grows linearly with $P$ while the interference power scales with the product of $P$ and the quantization error.
Since the quantization error is of the order $2^{-\frac{B_{BB}}{r-1}} = P^{-\frac{\alpha}{r-1}}$, the interference power scales as $P^{(1-\frac{\alpha}{r-1})}$ which gives a SINR that scales as $P^{\frac{\alpha}{r-1}}$.
Thus, the resulting DoF of each user is $\frac{\alpha}{r-1}$.
In order to obtain a sum DoF of $m (m \le K)$, the number of second-stage feedback bits should scale with SNR
\begin{eqnarray} \label{eq:feedbackscale}
B_{BB} = O\Big(m \frac{r - 1}{K}\frac{P_{dB}}{3} \Big).
\end{eqnarray}

\subsection{One-stage vs. Two-stage Feedback Scheme} \label{comp}
An analytical comparison between one-stage and two-stage feedback schemes is intractable due to intractability of a closed-form sum rate expression for the `TSF + Adp CB' scheme.
Nevertheless, we can get some insights based on the analysis and discussion in Sections \ref{onestage} and \ref{twostage}.
Specifically, we consider the following scenarios:
1. AoDs of each user are well separated;
2. very sparse channel (i.e., $L \rightarrow 1$);
3. dominant path (i.e., $|g_1| \gg |g_l|, \forall l \ge 2$);
4. very limited feedback overhead $B$.

From Proposition 1 and Remark 1, the multiuser interference can be effectively mitigated by the long-term channel statistics-based digital precoder in scenarios 1-3.
The per-user rate of the `OSF + Stat' scheme approaches the per-user interference-free rate.
Thus, the `OSF + Stat' scheme with $B_{RF} = B$ achieves higher rate than the `TSF + Adp CB' scheme with $B_{RF} = B - B_{BB}$.
In scenario 4, the splitting operation in `TSF + Adp CB' scheme results in very small $B_{RF}$ and $B_{BB}$.
From \eqref{eq:beam}, small $B_{RF}$ leads to coarse RF beamsteering, i.e., mismatch between the RF beamforming and the channel direction.
Additionally, small $B_{BB}$ incurs an inaccurate feedback of the effective channel knowledge because the quantization error is of the order $2^{-\frac{B_{BB}}{r-1}}$.
Thus, the `TSF + Adp CB' scheme with very limited $B$ yields an unfavourable rate performance.
In a nutshell, the `OSF + Stat' scheme outperforms the `TSF + Adp CB' counterpart in scenarios 1-4 in term of rate performance.
Meanwhile, the `OSF + Stat' scheme highly reduces the complexity of the signalling and feedback procedure by eliminating the second-stage feedback.

On the contrary, the `TSF + Adp CB' hybrid precoding scheme exceeds the `OSF + Stat' scheme in the regime of large $B$, where the `TSF + Adp CB' scheme with partial resources $B_{RF}$ is sufficient to provide precise first-stage beamforming.
The residual resources $B_{BB}$ gives an accurate channel quantization (i.e.,  $2^{-\frac{B_{BB}}{r-1}} \rightarrow 0$), resulting in efficient interference elimination.
By contrast, the `OSF + Stat' scheme using all feedback overhead $B$ for the first-stage RF beamsteering can only achieve a marginal RF beamforming gain over the `TSF + Adp CB' scheme using $B_{RF}$.
Meanwhile, the `OSF + Stat' scheme based on statistical CSIT-based digital precoder is less capable of mitigating interference.

In Section \ref{numerical}, a simulated comparison is provided to show the effectiveness of the `OSF + Stat' scheme over the `TSF + Adp CB' scheme in various aforementioned scenarios.
In order to fairly compare these two schemes, the optimal feedback allocation between the two stages of the `TSF + Adp CB' scheme is numerically computed\footnote{In general, the analytical computation of the optimum feedback allocation between two stages given certain total feedback overhead is not trivial and is beyond the scope of the present work.}.

%%%%%%%%%%%%%%%%%%%%%%%%%%%%%%%%%%%%%%%%%%%%%%%%%%%%%%%% update this in the thesis %%%%%%%%%%%%%%%%%%%%%%%%%%%%%%%%%%%%%%%%%%%%%%%%%%%%%%%%%%
%In Section \ref{numerical}, a simulated comparison is provided to show the efficacy of the `OSF + Stat' scheme over `TSF + Adp CB' in various aforementioned scenarios.
Moreover, we note that the beam search in the `OSF + Stat' scheme with $B_{RF} = B$ takes a bit longer than the `TSF + Adp CB' scheme with $B_{RF} = B - B_{BB}$.
Nevertheless, there is not a big gap between them when the total feedback is limited and the feedback allocation of the `TSF + Adp CB' scheme is optimized, as confirmed by the simulation results in Section \ref{numerical}.

\begin{figure}[t]
  \centering
  \includegraphics[width = 0.48\textwidth]{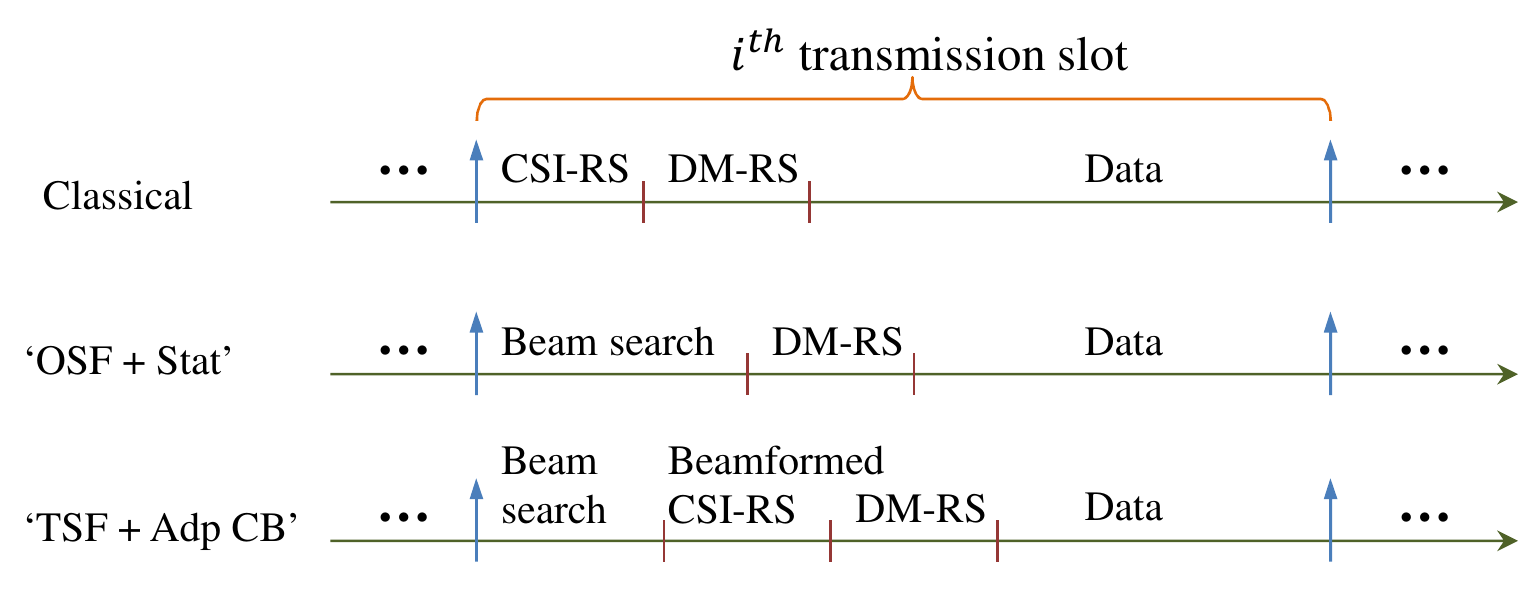}\\
  \caption{Illustration of signalling and feedback procedure for various schemes.} \label{fig_sfdiag}
\end{figure}

\subsection{Signalling and Feedback Protocol}
To operate the proposed hybrid precoding schemes, the signalling and feedback procedure is illustrated in Fig. \ref{fig_sfdiag} and described as follows.

Consider the classical microwave systems with $M$ transmit antennas and $N = M$ RF chains at the BS.
Using LTE-A framework \cite{Bocc2012}, channel state information reference signals (CSI-RS) are transmitted to enable the users to measure the instantaneous CSI, which is then fed back to the BS.
With this channel knowledge, the BS computes the precoders.
Then, the BS constructs the transmitted signals that are transmitted along demodulation reference signals (DM-RS) \cite{Lim2013} to enable the users to detect the desired signal.

In the mmWave systems with one-stage feedback scheme, we first search the beam that maximizes the desired signal of each user to overcome the severe pathloss of mmWave link.
Based on the analog beamformer and the channel covariance matrix, the BS designs the digital precoder and transmits the DM-RS.

In the mmWave systems with two-stage feedback scheme, beam search is first operated and the beam that maximizes the desired signal of each user is fed back to the BS.
Then, the BS transmits the beamformed CSI-RS to the users.
Each user estimates and reports the effective CSI to the BS.
Based on the analog beamformer and the quantized CSIT, the BS determines the digital precoder and then transmits the DM-RS.

%%%%%%%%%%%%%%%%%%%%%%%%%%%%%%%%%%%%%%%%%%%%%%%%%%%%%%%% update this in the thesis %%%%%%%%%%%%%%%%%%%%%%%%%%%%%%%%%%%%%%%%%%%%%%%%%%%%%%%%%%

\section{RATE SPLITTING WITH HYBRID PRECODING} \label{RS}
%\footnote{Imperfect CSIT refers to either imperfect channel quantization due to limited feedback or statistical CSIT. The second-order statistics can be viewed as a sort of imperfect knowledge of the instantaneous channel. Particularly, when the AoDs of users are mutually overlapped, statistical CSIT-based digital precoding is unable to completely eliminate the interference.}.

In previous section, the achievable rate of the conventional transmission strategy is highly degraded by multiuser interference due to either AoDs overlap in `OSF + Stat' scheme or limited feedback in `TSF + Adp CB' scheme.
In the context of multiuser mmWave systems, we introduce a rate splitting (RS) transmission strategy to tackle the residual interference.
RS enhances the rate performance and can be applied with both aforementioned hybrid precoding schemes.

With $B_{RF}$ for the first-stage analog beamforming, we consider the conventional transmission strategy (hereafter referred to as No-RS) with the `OSF + Stat' hybrid precoding as a baseline.
On one side, RS with the `OSF + Stat' hybrid precoding achieves rate gain over the baseline owing to the benefits of RS.
On the other side, by using the same $B_{RF}$ for the first-stage analog beamforming and extra feedback $B_{BB}$ for the second-stage channel quantization, the conventional No-RS with the `TSF + Adp CB' hybrid precoding also enables rate gain over the baseline.
In this section, we mainly investigate how much second-stage feedback can be saved by implementing RS.
%\footnote{Normally, one can incorporate RS with `TSF + Adp CB' to enable rate gain over the conventional multiuser transmission with `TSF + Adp CB' of Section III-B. In this work, nevertheless, we are primarily interested in the comparison between RS + `OSF + Stat' with less feedback and No-RS + `TSF + Adp CB' with more feedback.}.

The RS transmission strategy superposes a common message on top of all users' private messages.
Thus, the conventional No-RS strategy is a sub-scheme of RS.
Compared with \eqref{eq:tx_sigbc}, the transmitted signal of RS can be written as
\begin{eqnarray} \label{eq:tx_sigrs}
\mathbf{x} = \sqrt{P_{c}}\, \mathbf{F}\, \mathbf{w}_{c} s_{c} + \sum^K_{k = 1}  \sqrt{P_{k}}\, \mathbf{F} \,\mathbf{w}_{k} \, s_{k},
\end{eqnarray}
where $\mathbf{w}_{c}$ is the precoding vector of the common message $s_c$.
In \eqref{eq:rx_sigbc}, the private message transmissions are interference-limited at high SNR.
The basic idea of RS is to transmit the private messages with a fraction of the total power such that the private messages are decoded in the non-interference-limited SNR regime.
In addition, a common message is transmitted using the remaining power which gives rise to a rate enhancement \cite{Dai2016}.

In line with the conventional No-RS strategy, the RF beamformer $\mathbf{F}$ and the digital precoder $\mathbf{w}_k$ are designed as in Section \ref{onestage} while uniform power allocation is performed on the private messages.
We mainly focus on the power splitting between the common and private messages and the precoder design of the common message.
A fraction $t \in (0, \, 1]$ of the total power is uniformly allocated to the private messages while the remaining power is given to the common message, i.e., $P_k = Pt/K$ and $P_c = P(1 - t)$.
At the user side, each user decodes first the common message by treating all private messages as noise.
After removing the decoded common message by SIC, each user decodes its own private message.
Thus, the SINRs of the common message and the private message experienced by user $k$ are written as
\begin{eqnarray} \label{eq:rs_sinr1}
\text{SINR}_{k}^{c} &=& \frac{P_{c} \, |\mathbf{h}^H_{k} \mathbf{F} \mathbf{w}_{c} |^2 }{1 + \sum_{j=1}^K P_{j} \, |\mathbf{h}^H_{k} \mathbf{F} \mathbf{w}_{j} |^2},  \\
\text{SINR}_{k}^{p} &=& \frac{P_{k} \, |\mathbf{h}^H_{k} \mathbf{F} \mathbf{w}_{k} |^2 }{1 + \sum_{j \ne k} P_{j} \, |\mathbf{h}^H_{k} \mathbf{F} \mathbf{w}_{j} |^2}.
\end{eqnarray}
The achievable rate of the common message is given as
\begin{eqnarray} \label{eq:rs_ratecom}
R^{c} =  \underset{k}{\min} \; \{R^c_k\} = \underset{k}{\min} \; \big\{ \log_2 (1 + \text{SINR}^{c}_k) \big\},
\end{eqnarray}
which guarantees that the common message can be successfully decoded by all users.
The sum rate of the private messages is given as $R^{p} = \sum^{K}_{k=1} R^p_{k} = \sum^{K}_{k=1} \log_2 (1 + \text{SINR}^p_{k})$.
Then, the sum rate of RS is $R^{RS}_{\scriptstyle{\text{sum}}} = R^{c} + R^{p}$.
In order to properly design $t$ and $\mathbf{w}_c$, we need to derive the sum rate expression of RS.

$\textbf{Proposition 3:}$ The average sum rate of RS with the `OSF + Stat' hybrid precoding is lower bounded as
\begin{eqnarray} \label{eq:achraters}
\!\!\!\!\!\mathbb{E} (R^{RS}_{\text{sum}}) \!\!\!\!&\ge&\!\!\!\!  \underset{k}{\text{min}} \bigg\{ \log_2 \bigg(1 + \frac{e^{-\gamma}\cdot P(1-t)\, \mathbf{w}^H_c \mathbf{R}_{k,\text{eff}} \mathbf{w}_c}{1 +\frac{Pt}{K} \sum^K_{j=1} \mathbf{w}^H_j \mathbf{R}_{k,\text{eff}} \mathbf{w}_j} \bigg) \bigg\} \nonumber \\
&+& \!\!\sum^K_{k = 1} \log_2 \bigg(1 + \frac{e^{-\gamma} \cdot \frac{Pt}{K} \mathbf{w}^H_k \mathbf{R}_{k,\text{eff}} \mathbf{w}_k }{1 + \frac{Pt}{K} \sum_{j \ne k} \mathbf{w}^H_j \mathbf{R}_{k,\text{eff}} \mathbf{w}_j} \bigg).
\end{eqnarray}
\begin{IEEEproof}
See Appendix \ref{sec:prop3}.
\end{IEEEproof}

According to \eqref{eq:achraters}, the precoder ($\mathbf{w}_c$) and the power splitting ratio ($t$) can be designed.
\subsection{Precoder design}
%\emph{1) Precoder design}:
With $\mathbf{R}_{k,\text{eff}} $ known at the BS and $\mathbf{w}_j$ designed as \eqref{eq:GE}, $\mathbf{w}_c$ can be optimized by solving the following max-min problem subject to a power constraint
\begin{eqnarray} \label{eq:precoderopt}
\mathcal{P}1: && \underset{\mathbf{w}_{c}}{\text{max}} \;\; \underset{k}{\text{min}} \;\, \frac{1}{\beta_k} \mathbf{w}^H_c \mathbf{R}_{k,\text{eff}} \mathbf{w}_c  \\
&& \text{s.t.} \quad \mathbf{w}^H_{c} \, \mathbf{M} \, \mathbf{w}_c \le 1, \label{eq:precoderopt22}
\end{eqnarray}
where $\mathbf{M} = \mathbf{F}^H \mathbf{F}$ and $\beta_k = 1 + \frac{Pt}{K} \sum^K_{j=1} \mathbf{w}^H_j \mathbf{R}_{k,\text{eff}} \mathbf{w}_j$.
Note that the constraint \eqref{eq:precoderopt22} in problem $\mathcal{P}1$ should be met with equality at an optimum, otherwise $\mathbf{w}_c$ can be scaled up, thereby improving the objective and contradicting optimality.
Thus, the optimized $\mathbf{w}_c$ still makes the power constraint hold.
Since $\mathbf{M}$ is positive semi-definite, \eqref{eq:precoderopt22} is a convex constraint.
By replacing $\mathbf{w}_c$ with $\mathbf{x}$ and introducing a slack variable, $\mathcal{P}1$ is equivalently reformulated as
\begin{eqnarray} \label{eq:precoderopt2}
\hspace{-0.5cm} \mathcal{P}2: && \underset{\mathbf{x},\, t}{\text{max}} \;\; t \\
&& \text{s.t.} \quad \mathbf{x}^H \mathbf{R}_{k,\text{eff}} \mathbf{x} \ge \beta_k t, \, \forall k \label{pd2} \\
&& \qquad \, \mathbf{x}^H \, \mathbf{M} \, \mathbf{x} \le 1, \nonumber
\end{eqnarray}
which is known as NP-hard problem due to non-convex constraint \eqref{pd2}. A well-known conservative solution is semi-definite relaxation (SDR) followed by semi-definite programming (SDP) and Gaussian randomization \cite{Tom2006}.
Rather, we cope with $\mathcal{P}2$ by successive convex approximation (SCA) approach \cite{Meh2015} for its lower worse-case complexity.
In addition, SCA converges to a KKT stationary point for the original problem $\mathcal{P}1$ \cite{Beck2010}.
For any $\mathbf{z}$ and PSD $\mathbf{R}_{k,\text{eff}} $, we have $(\mathbf{x} - \mathbf{z})^H \mathbf{R}_{k,\text{eff}} \,(\mathbf{x} - \mathbf{z}) \ge 0$ and therefore
\begin{eqnarray} \label{eq:sca1}
\mathbf{x}^H \mathbf{R}_{k,\text{eff}} \, \mathbf{x} \ge 2 \,\text{Re}\{\mathbf{z}^H \mathbf{R}_{k,\text{eff}} \,\mathbf{x}\} - \mathbf{z}^H \mathbf{R}_{k,\text{eff}}\, \mathbf{z}.
\end{eqnarray}

Substituting the convex constraint of \eqref{eq:sca1} into \eqref{pd2} leads to the following convex problem
\begin{eqnarray} \label{eq:precoderopt3}
\mathcal{P}3: \!\!\!\!\!&& \underset{\mathbf{x}, \,t}{\text{max}} \;\; t \nonumber \\
&& \text{s.t.} \, 2 \,\text{Re}\{\mathbf{z}^H \mathbf{R}_{k,\text{eff}} \,\mathbf{x}\} - \mathbf{z}^H \mathbf{R}_{k,\text{eff}}\, \mathbf{z} \ge \beta_k t, \, \forall k \\
&& \qquad \; \mathbf{x}^H \, \mathbf{M} \, \mathbf{x} \le 1, \nonumber
\end{eqnarray}
which can be easily formulated as a second-order cone programming (SOCP) problem. The optimal solution $\mathbf{x}^{\star}$ can be efficiently obtained for given $\mathbf{z}$.
Then, $\mathbf{z}$ is iteratively updated by $\mathbf{z} = \mathbf{x}^{\star}$ and used in the next iteration, yielding a sequence of feasible solutions with non-increasing objective values.
%%%%%%%%%%%%%%%%%%%%%%%%%%%%%%%%%%%%%%%%%%%%%%%%%%%%%%%%%%%%%%%%%%%%%%%%%%%%%%%%%%%%%%%%%%%%%%%%%%
The SCA approach needs a feasible point for initialization, which is difficult to obtain in general.
However, SDR + randomization can easily find a feasible solution to the original problem, which can be used as a good initialization for SCA.
Alternatively, the feasible point pursuit method proposed by \cite{Meh2015} can also be used to find a feasible initial point.
%%%%%%%%%%%%%%%%%%%%%%%%%%%%%%%%%%%%%%%%%%%%%%%%%%%%%%%%%%%%%%%%%%%%%%%%%%%%%%%%%%%%%%%%%%%%%%%%%%
The SCA algorithm is summarized in Algorithm \ref{hyalg2}.

\begin{algorithm}[t]
\caption{: SCA Algorithm}\label{hyalg2}
\begin{algorithmic}[1]
\State \textbf{Initialization}: set $i = 0$ and generate an initial point $\mathbf{z}_0$
\State \textbf{Repeat}
\State \quad solve \eqref{eq:precoderopt3} with $\mathbf{z} = \mathbf{z}_i$ and denote $\mathbf{x}^{\star}$ as the solution
\State \quad update $\mathbf{z}_{i + 1} = \mathbf{x}^{\star}$
\State \quad set $i = i + 1$
\State \textbf{Until convergence}
\end{algorithmic}
\end{algorithm}

\begin{figure*}[t] \centering
\subfigure[]{ \label{fig1} \includegraphics[width = 0.45\textwidth]{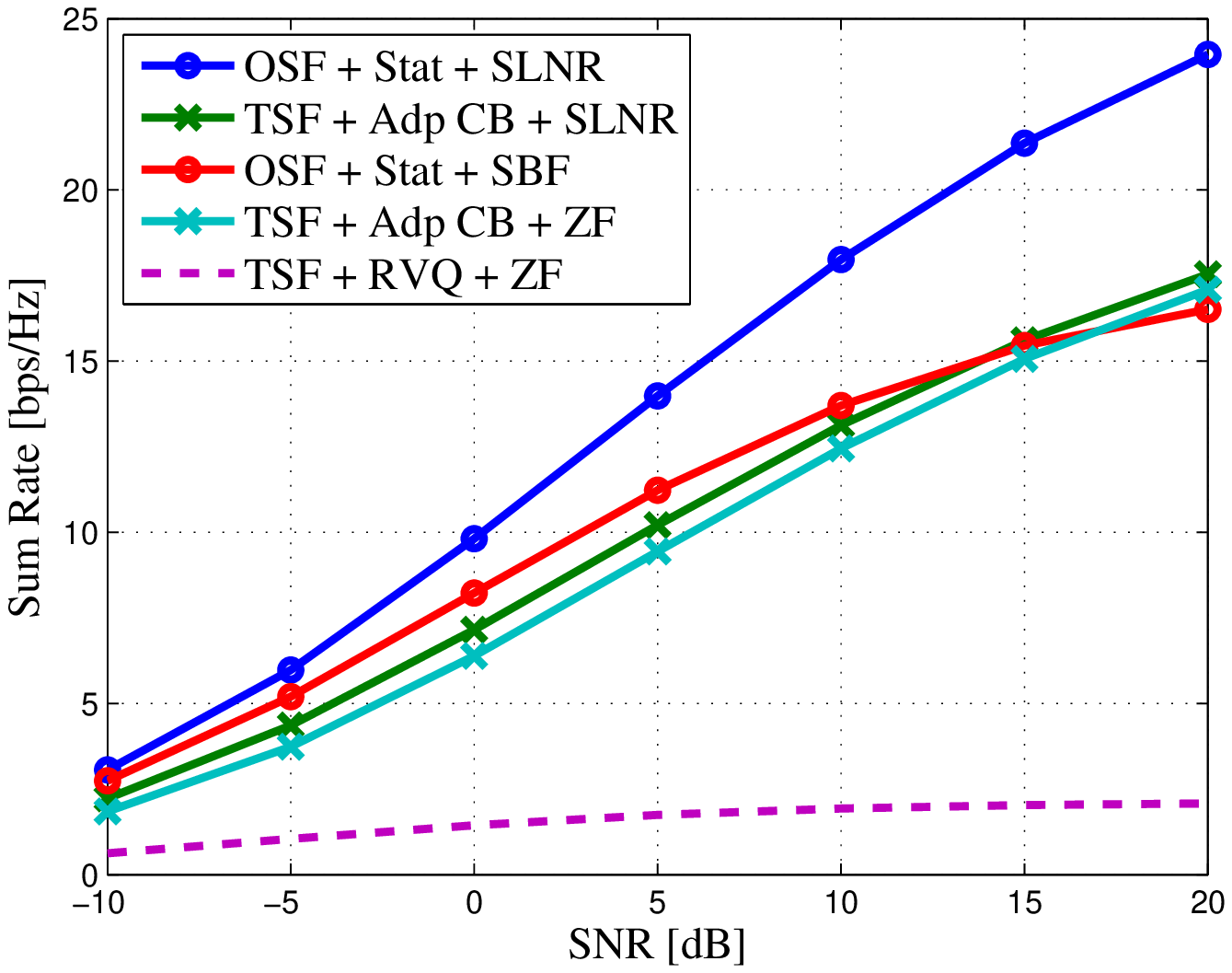}  } \hspace{-5pt}
\subfigure[]{ \label{fig2} \includegraphics[width = 0.45\textwidth]{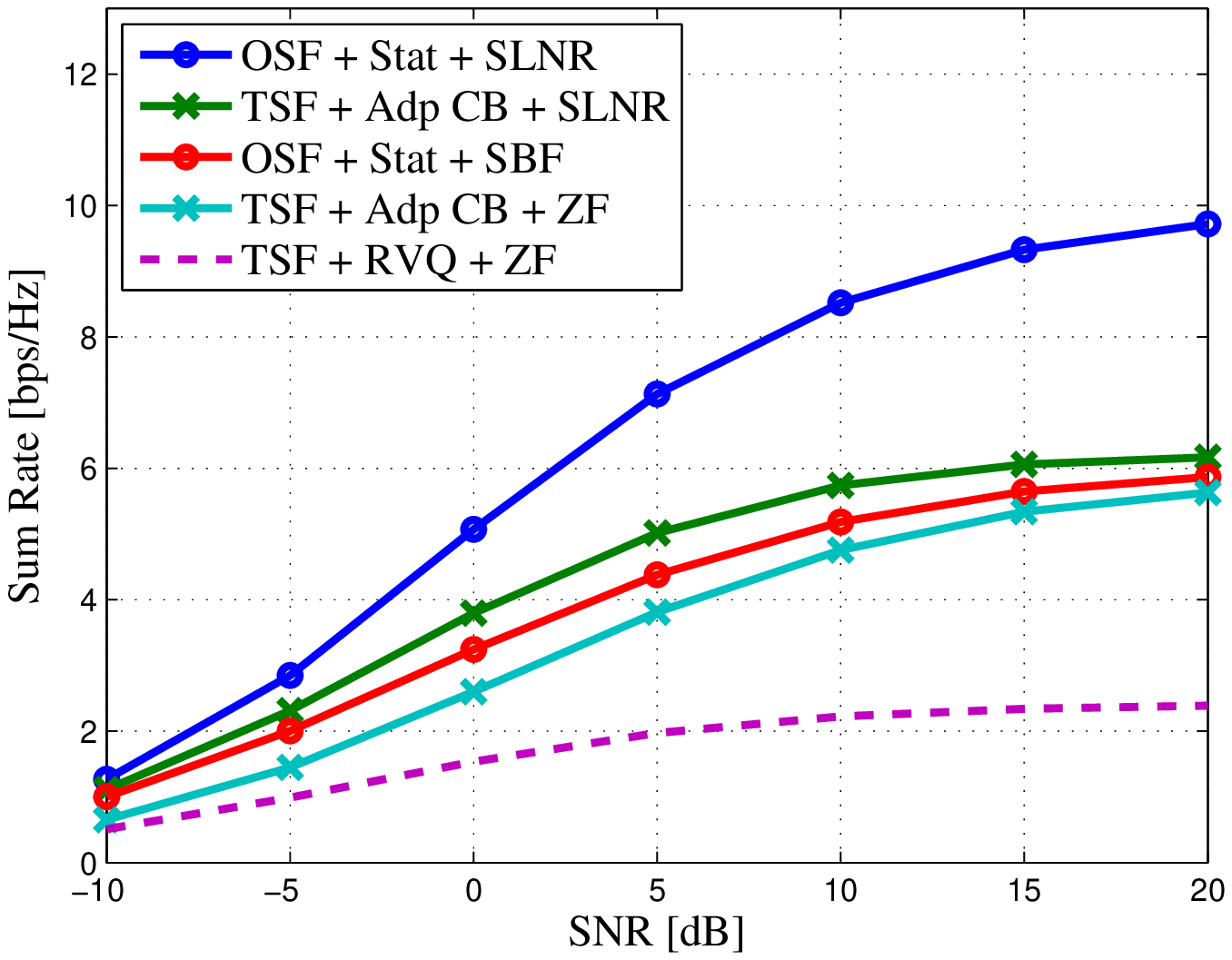} }
\caption{Comparison between the hybrid precoding schemes under various channel sparsity and $B = 6$; (a) $L = 2$, (b) $L = 15$.} \label{fig12}
\end{figure*}

\subsection{Power allocation design}
%\emph{2) Power allocation design}:
The optimal power splitting ratio $t$ can be determined by maximizing the lower bound on average sum rate \eqref{eq:achraters} with line search.
By contrast, we compute a suboptimal but effective and insightful power allocation.
Recall that the common message in RS is dedicated to overcome multiuser interference and its achievable rate is subject to a minimum constraint \eqref{eq:rs_ratecom}.
In the low/non-interference-limited SNR regime, private messages transmission works well and therefore the total power is allocated to the private messages, i.e., $t = 1$.
In this case, RS turns into conventional No-RS.

At high SNR, exploiting full power to transmit the private messages only offers marginal sum rate gain by virtue of multiuser interference.
Thus, only a fraction of the total power is assigned to the private messages such that the private message decoding can be placed back to the non-interference-limited SNR regime.
The basic idea of power allocation for RS is to compute the saturation point of the achievable rate of private message, where the multiuser interference becomes the dominant factor for the performance.
Any power beyond is reserved for the common message.
Specifically, in \eqref{eq:achraters}, the interference term of the private message $\Upsilon \triangleq \frac{Pt}{K} \sum_{j \ne k} \mathbf{w}^H_j \mathbf{R}_{k,\text{eff}} \mathbf{w}_j$ and the noise power is 1.
The saturation point lies in the regime $\Upsilon > 1$ and can be computed by letting $\Upsilon = K$ for simplicity\footnote{Apparently, the value of $\Upsilon$ satisfying $\Upsilon > 1$ is not unique. Nevertheless, the effectiveness of choosing $\Upsilon = K$ has been demonstrated by \cite{Dai2016}.}.
Overall, the power splitting ratio between the common and the private messages for arbitrary SNR is therefore designed as
\begin{eqnarray} \label{eq:powerallc}
t = \text{min} \; \Big\{ \frac{K}{P \Gamma}, \; 1 \Big\},
\end{eqnarray}
where $\Gamma = \underset{k}{\text{min}} \, \big\{ \frac{1}{K} \sum_{j \ne k} \mathbf{w}^H_j \mathbf{R}_{k,\text{eff}} \mathbf{w}_j \big\}$.
The minimum is imposed to guarantee that $\frac{Pt}{K} \sum_{j \ne k} \mathbf{w}^H_j \mathbf{R}_{k,\text{eff}} \mathbf{w}_j \gg 1$ for $\forall k$.
The overall RS transmission strategy is presented in Algorithm \ref{hyalg3}.
First, the base stations determines the analog beamformer based on beam search and the digital precoder based on channel covariance matrix. Then, the BS computes the power splitting ratio $t$ according to the precoder design and system parameters. If $t = 1$, the BS transmits using conventional multiuser hybrid precoding scheme. Otherwise, the BS needs to further calculate the precoder for the common message and transmits using the RS scheme with hybrid precoding.

\begin{algorithm}[t]
\caption{: Overall RS transmission with hybrid precoding}\label{hyalg3}
\begin{algorithmic}[1]
\State Base station: sets $\mathbf{F}$ from \eqref{eq:beam} and $\mathbf{R}_{k,\text{eff}} = \mathbf{F}^H \mathbf{R}_{k} \mathbf{F}$
\State \qquad \qquad  \quad \, computes $\mathbf{w}_k$ from \eqref{eq:GE} and $t$ from \eqref{eq:powerallc}
\State If $t < 1$
\State \qquad Base station: determines $\mathbf{w}_c$ as the solution of $\mathcal{P}1$
\State \qquad \qquad\qquad \quad \, constructs the TX signal as \eqref{eq:tx_sigrs}
\State \qquad Users: decode the common message $\rightarrow$ SIC $\rightarrow$
\State \qquad \qquad \quad $\rightarrow$ their own private messages
\State Else
\State \qquad Base station: constructs the TX signal as \eqref{eq:tx_sigbc}
\State \qquad Users: decode directly their own private messages
\end{algorithmic}
\end{algorithm}

\emph{Remark 2}: In general scenarios such as partially overlapped AoDs between users, the sum rate of the baseline (i.e., conventional No-RS strategy with the `OSF + Stat' hybrid precoding scheme) is interference-limited at high SNR.
By contrast, in the `OSF + Stat' precoded RS strategy, the power allocation \eqref{eq:powerallc} guarantees that the private message transmission of RS achieves almost the same sum rate as the baseline.
Meanwhile, the common message transmission of RS with the remaining power leads to rate enhancement which can be quantified as $R_c$ \eqref{eq:rs_ratecom}.
Therefore, the sum DoF gain of the `OSF + Stat' precoded RS over the baseline approaches 1.

Moreover, the `TSF + Adp CB' precoded conventional No-RS strategy also enables better performance than the baseline by virtue of extra second-stage feedback.
Based on \eqref{eq:feedbackscale}, `TSF + Adp CB' requires extra feedback scaling as $B_{BB} = O(\frac{r - 1}{K}\frac{P_{dB}}{3})$ to achieve a sum DoF of 1.
As the rank $r$ of the effective channel covariance matrix $\mathbf{R}_{k,\text{eff}} = \mathbf{F}^H \mathbf{R}_{k} \mathbf{F} \in \mathbb{C}^{K \times K}$ is constrained to the channel sparsity $L$, we have $r = \min(L, K)$. Consequently, the `OSF + Stat' precoded RS can achieve the same sum DoF as the `TSF + Adp CB' precoded conventional No-RS that is driven by extra second-stage feedback\footnote{It is worth noting that simply doing TDMA achieves DoF of 1. However, the rate gain of RS over TDMA has been demonstrated in \cite{Clerckx2016, Hao2015, Dai2016}.}.
It implies that RS enables significant saving both in the downlink training and in the CSIT uplink feedback.

\section{SIMULATION RESULTS} \label{numerical}
Numerical results are provided to validate the effectiveness of the proposed hybrid precoding schemes and RS strategy.
We consider a system model with ULA equipped at the BS and a channel model with AoDs uniformly distributed in $[0,\, \pi]$.
The codebook $\mathcal{C}_{\text{iid}}$ is designed using Grassmannian line packing.

\subsection{Multiuser Hybrid Precoding} \label{Numerical_MU}
This section simulates the conventional No-RS transmission strategy in a setup with $M = 64, K = 4, B = 6$.
In Fig. \ref{fig12}, we evaluate the performance of the proposed one-stage feedback plus channel statistics-based (`OSF + Stat') and two-stage feedback plus adaptive codebook-based (`TSF + Adp CB') hybrid precoding schemes.
The `OSF + Stat' scheme uses all the feedback bits for the first-stage analog beamforming, i.e., $B_{RF} = B, B_{BB} = 0$.
By contrast, in `TSF + Adp CB' scheme, the total feedback overhead $B$ is divided into the first-stage analog beamsteering ($B_{RF}$) and the second-stage channel quantization ($B_{BB}$).
The optimal feedback allocation is numerically computed and the corresponding sum rate of `TSF + Adp CB' is plotted.
The effectiveness of the `OSF + Stat' scheme is verified by comparing with the `optimized' `TSF + Adp CB' scheme.

Both SLNR-based and SBF/ZF digital precoder designs are considered.
The \textbf{baseline} (`TSF + RVQ + ZF' \cite{Alk2015}) is plotted where RVQ codebook is used to quantize the effective channel and the BS designs the digital precoder as ZF based on the quantized channels.
Fig. \ref{fig12} shows that the proposed hybrid precoding schemes largely outperform the baseline due to the exploitation of the channel statistics.

\begin{figure*}[t] \centering
\subfigure[]{ \label{fig32} \includegraphics[width = 0.45\textwidth]{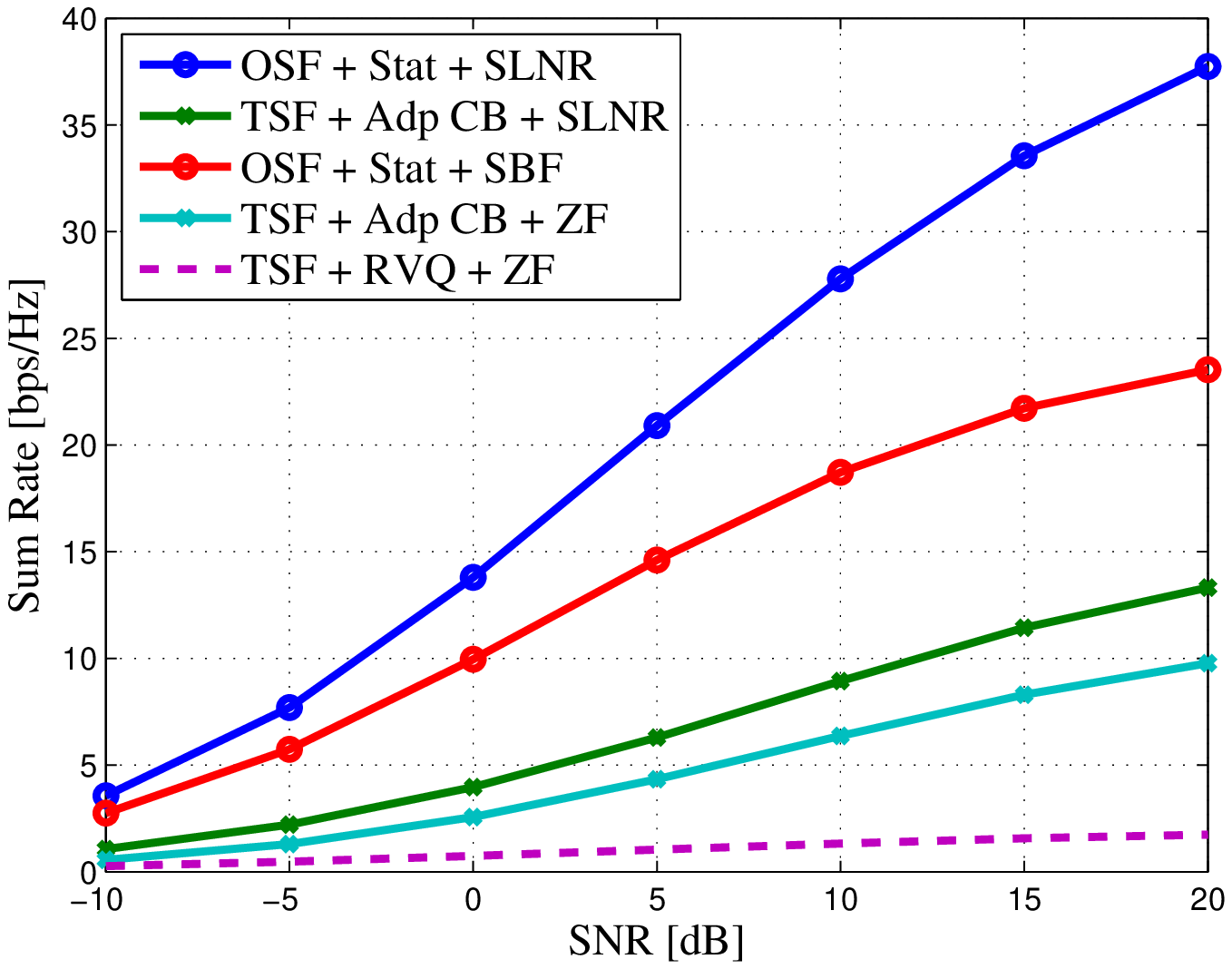}  } \hspace{-5pt}
\subfigure[]{ \label{fig42} \includegraphics[width = 0.45\textwidth]{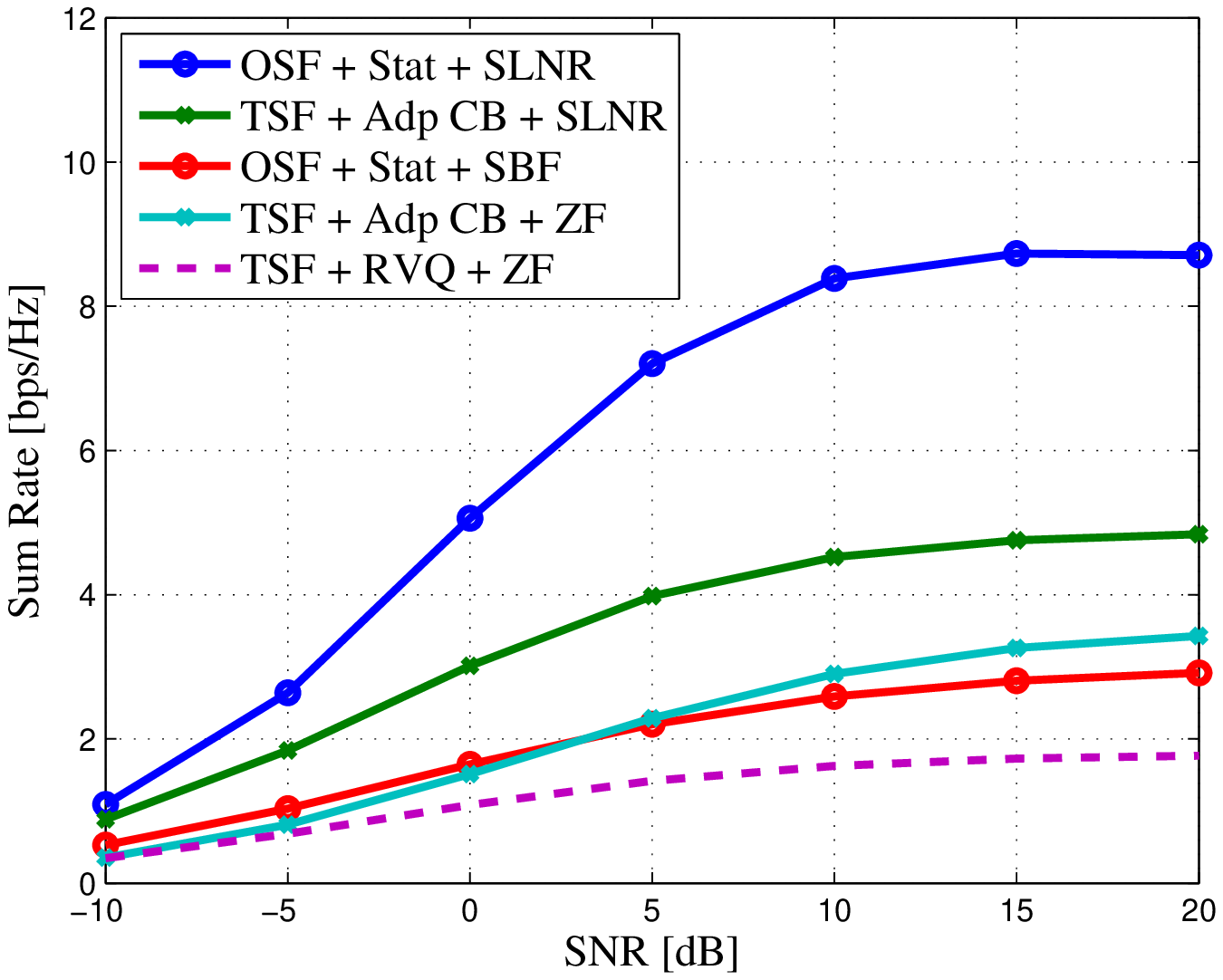} }
\caption{Comparison between the hybrid precoding schemes under various channel sparsity and $B = 6, K = 8$; (a) $L = 2$, (b) $L = 15$.} \label{fig342}
\end{figure*}
\begin{figure*}[htbp] \centering
\subfigure[]{ \label{fig3} \includegraphics[width = 0.45\textwidth]{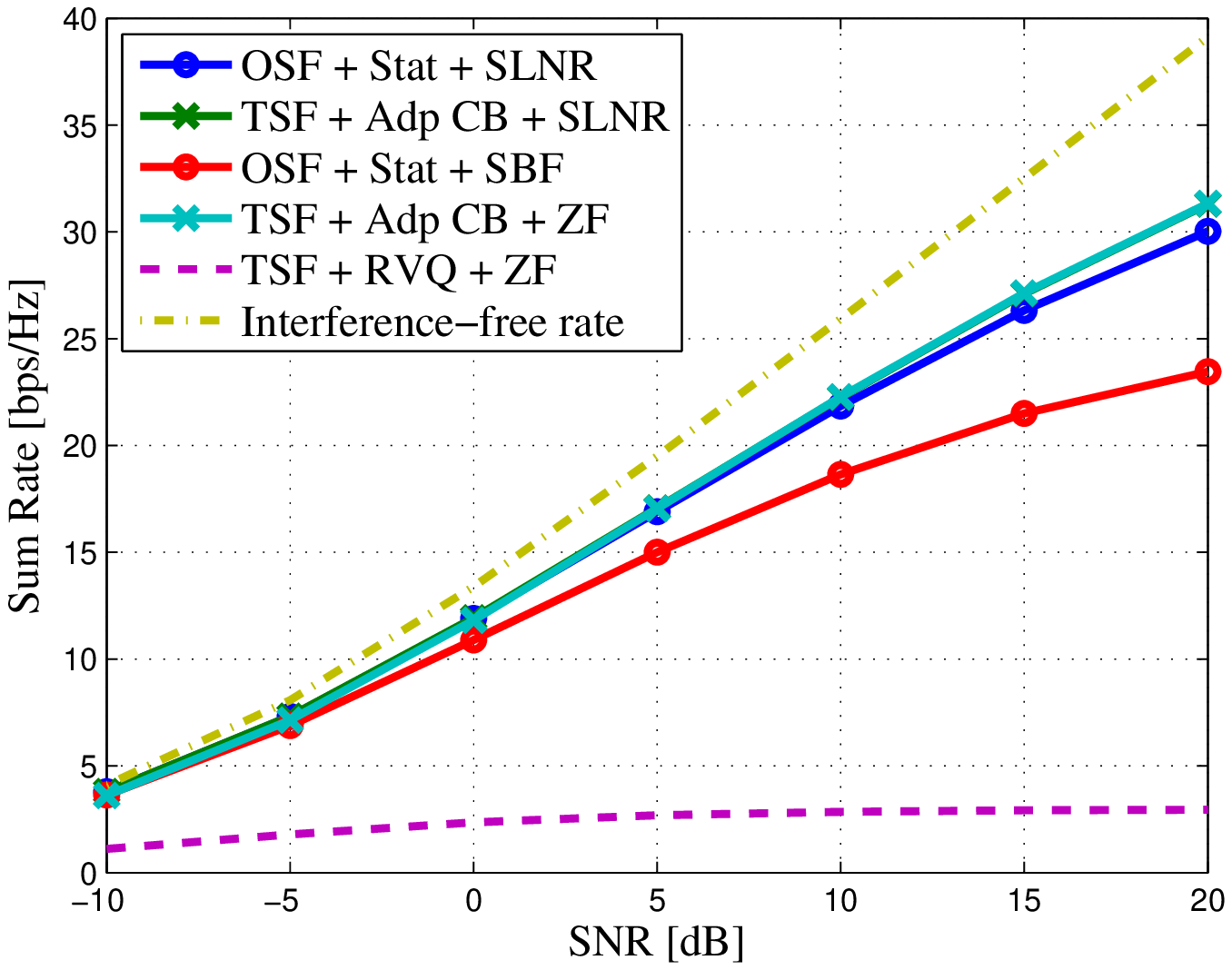}  } \hspace{-5pt}
\subfigure[]{ \label{fig4} \includegraphics[width = 0.45\textwidth]{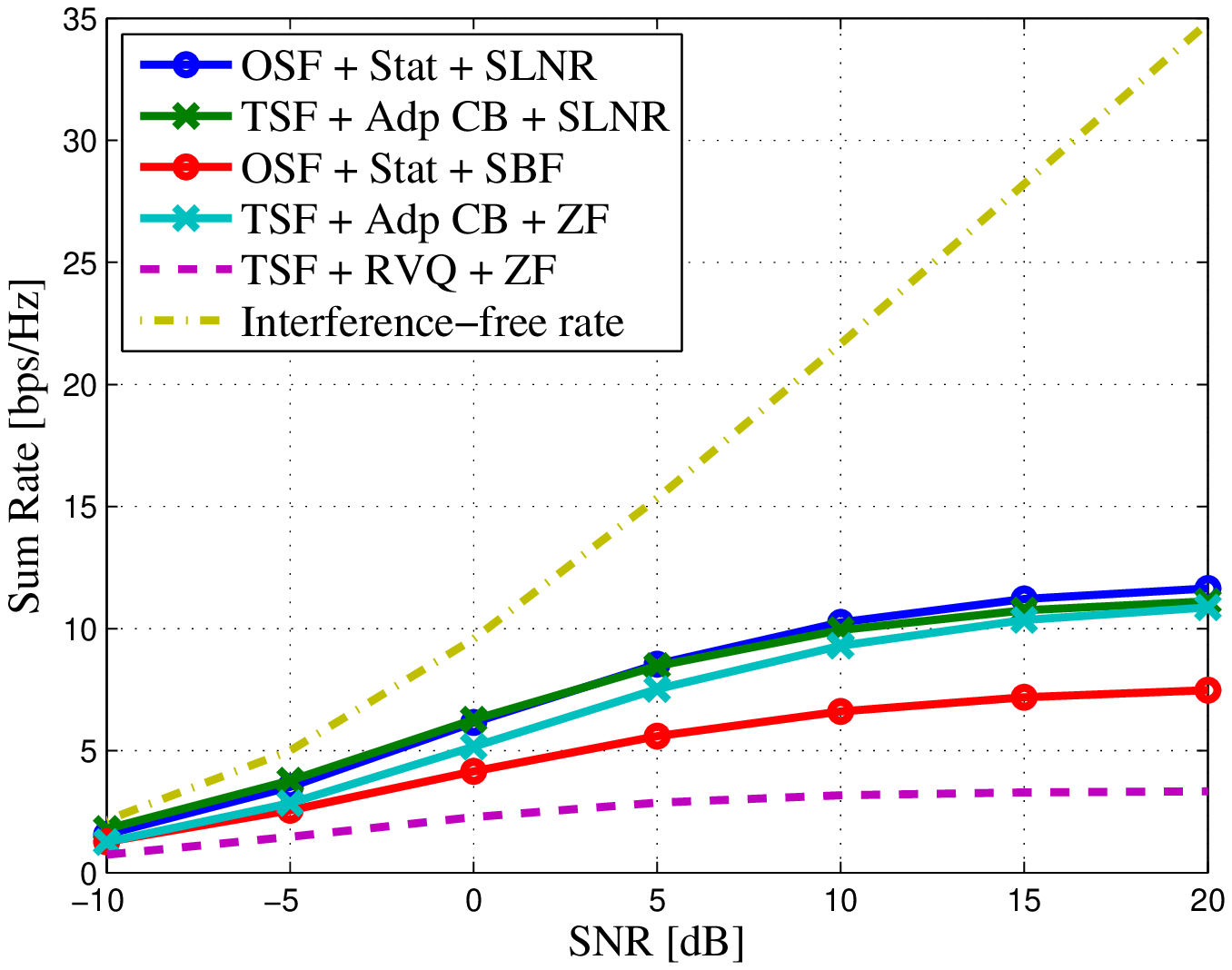} }
\caption{Comparison between the hybrid precoding schemes under various channel sparsity and $B = 10$; (a) $L = 2$, (b) $L = 15$.} \label{fig34}
\end{figure*}

It implies that adaptive codebook captures the characteristics of the mmWave channel better than RVQ.
Moreover, SLNR-based digital precoder design that maximizes a lower bound on the averaged SLNR enables higher rate than ZF, since it takes into account both the desired signal and the interference.

\begin{figure*}[t] \centering
\subfigure[]{ \label{fig5} \includegraphics[width = 0.45\textwidth]{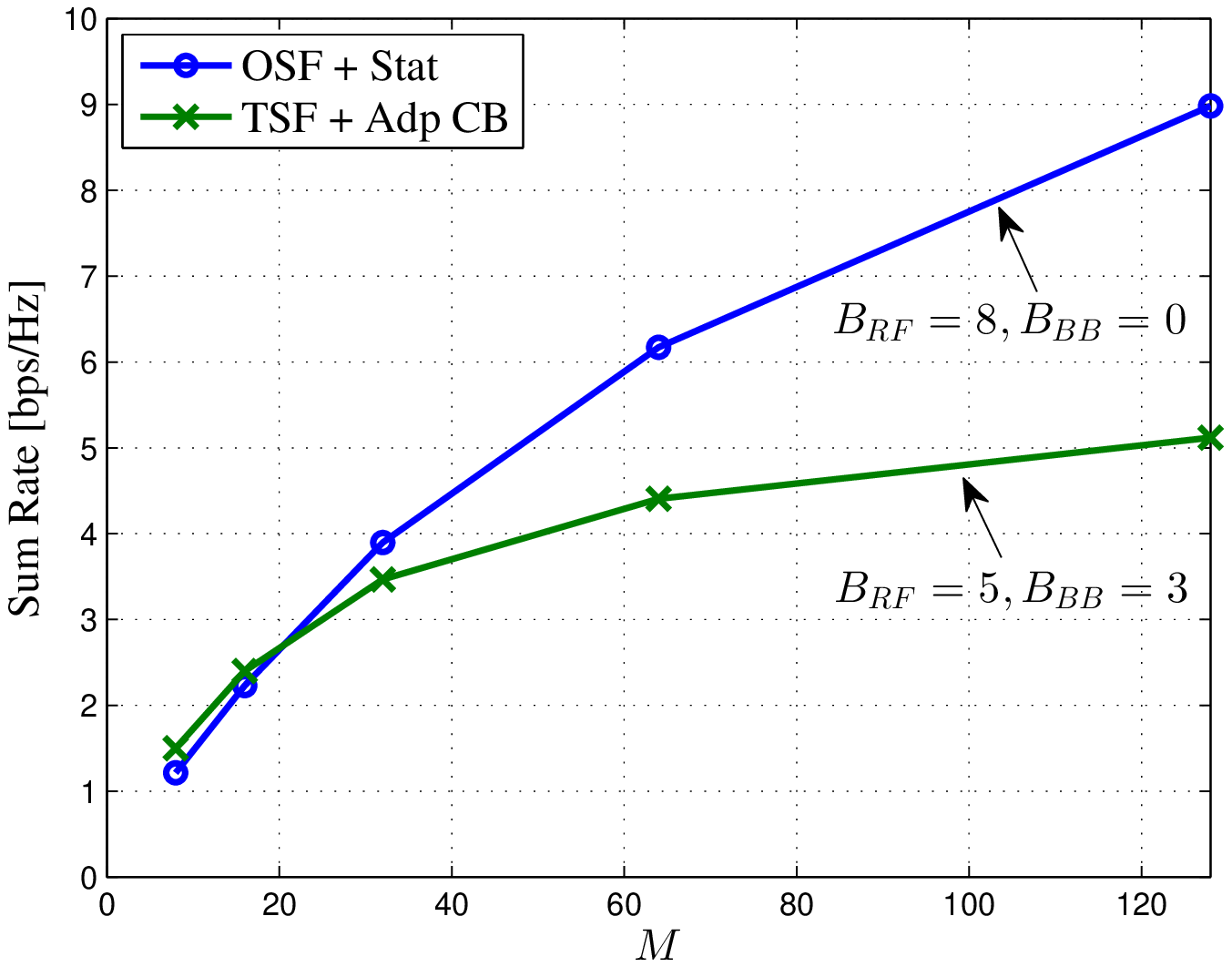}  } \hspace{-5pt}
\subfigure[]{ \label{fig6} \includegraphics[width = 0.45\textwidth]{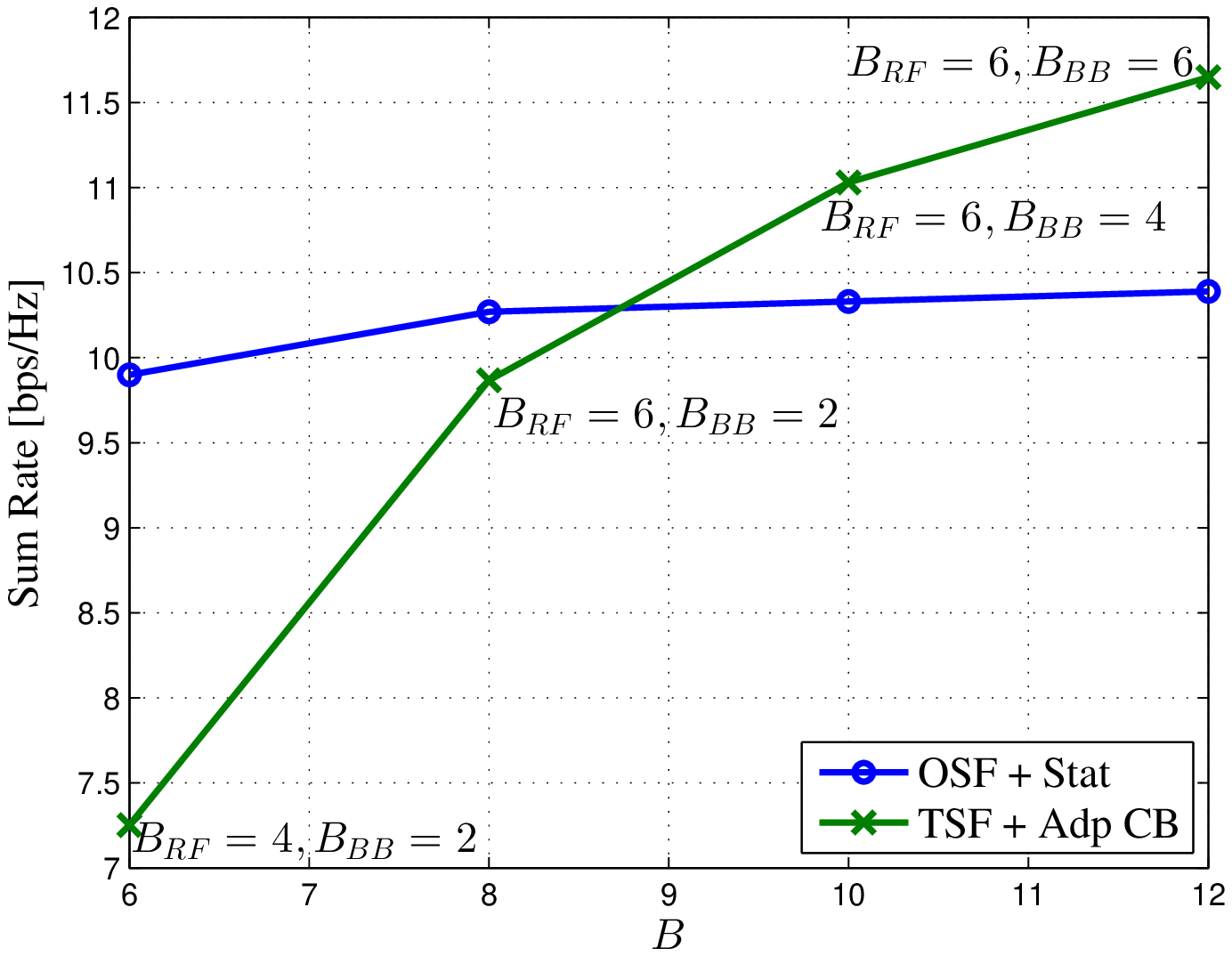} }
\caption{Comparison between the hybrid precoding schemes with various $M$ and $B$; (a) $B = 8$, SNR $= -5$ dB,  (b) $M = 32$, SNR $= 5$ dB.} \label{fig56}
\end{figure*}

The optimal splitting of the total feedback overhead $B = 6$ in `TSF + Adp CB' scheme is numerically obtained as $B_{RF} = 5, B_{BB} = 1$.
In Fig. \ref{fig1}, it can be seen that the proposed `OSF + Stat + SLNR' scheme achieves much higher sum rate than the `TSF + Adp CB' scheme.
For instance, the gain of `OSF + Stat + SLNR' over `TSF + Adp CB + SLNR' is 4.85 bps/Hz (equivalently, 37\% improvement) at SNR = 10 dB.
The `OSF + Stat' scheme utilizes all feedback resources to design precise RF beamforming.
Meanwhile, the channel covariance matrix is low rank due to channel sparsity ($L = 2$) and leaves nullspace to efficiently mitigate multiuser interference.
In Fig. \ref{fig2} where $L = 15$, the rate performance of the `OSF + Stat' scheme still exceeds the `TSF + Adp CB' scheme.
For instance, the gain of `OSF + Stat + SLNR' over `TSF + Adp CB + SLNR' is 2.8 bps/Hz (equivalently 48\% improvement) at SNR = 10 dB.
In `TSF + Adp CB' scheme, the optimal splitting of total feedback $B = 6$ is numerically obtained as $B_{RF} = 4, B_{BB} = 2$.
When a limited amount of feedback bits $B = 6$ is attainable, the splitting operation in `TSF + Adp CB' scheme leads to coarse analog beamsteering and inaccurate channel quantization.
Therefore, these observations in Fig. \ref{fig12} verify the comments drawn in Section \ref{comp}.
Namely, the proposed `OSF + Stat' scheme exhibits substantial sum rate gain over various baselines for very limited feedback system/very sparse channels.
Fig. \ref{fig342} has the same setup as Fig. \ref{fig12} but with a larger number of users $K = 8$.
Similar observations have been noted.

In Fig. \ref{fig34}, we consider the same setup as Fig. \ref{fig12} but with a larger feedback overhead $B = 10$.
The sum of interference-free per-user rate $R_{\text{sum}} = \sum^K_{k=1} R^s_k$, where $R^s_k$ is given as \eqref{eq:achrate1}, is plotted as an upper bound on the sum rate.
In `TSF + Adp CB' scheme, the optimal splitting of total feedback $B = 10$ is numerically obtained as $B_{RF} = 8, B_{BB} = 2$ in Fig. \ref{fig3} and $B_{RF} = 7, B_{BB} = 3$ in Fig. \ref{fig4}.
Fig. \ref{fig3} shows that the proposed `OSF + Stat + SLNR' scheme obtains a rate performance similar to that of the the `optimized' `TSF + Adp CB' scheme.
However, the `OSF + Stat + SLNR' scheme does not require the downlink training and uplink feedback of the effective channel and therefore is still preferable to the `TSF + Adp CB' scheme.
It can be concluded that multiuser interference is effectively mitigated in `OSF + Stat' scheme (due to low rank channel covariance matrix) and in `TSF + Adp CB' scheme (owing to a relatively large number of feedback).
%The gap between the per-user interference-free rate and the achievable rate of the `OSF + Stat + SLNR' precoded BC scheme enlarges as the SNR increases.
%This behavior matches the observation in Corollary 1.

\begin{figure*}[t] \centering
\subfigure[]{ \label{fig7} \includegraphics[width = 0.45\textwidth]{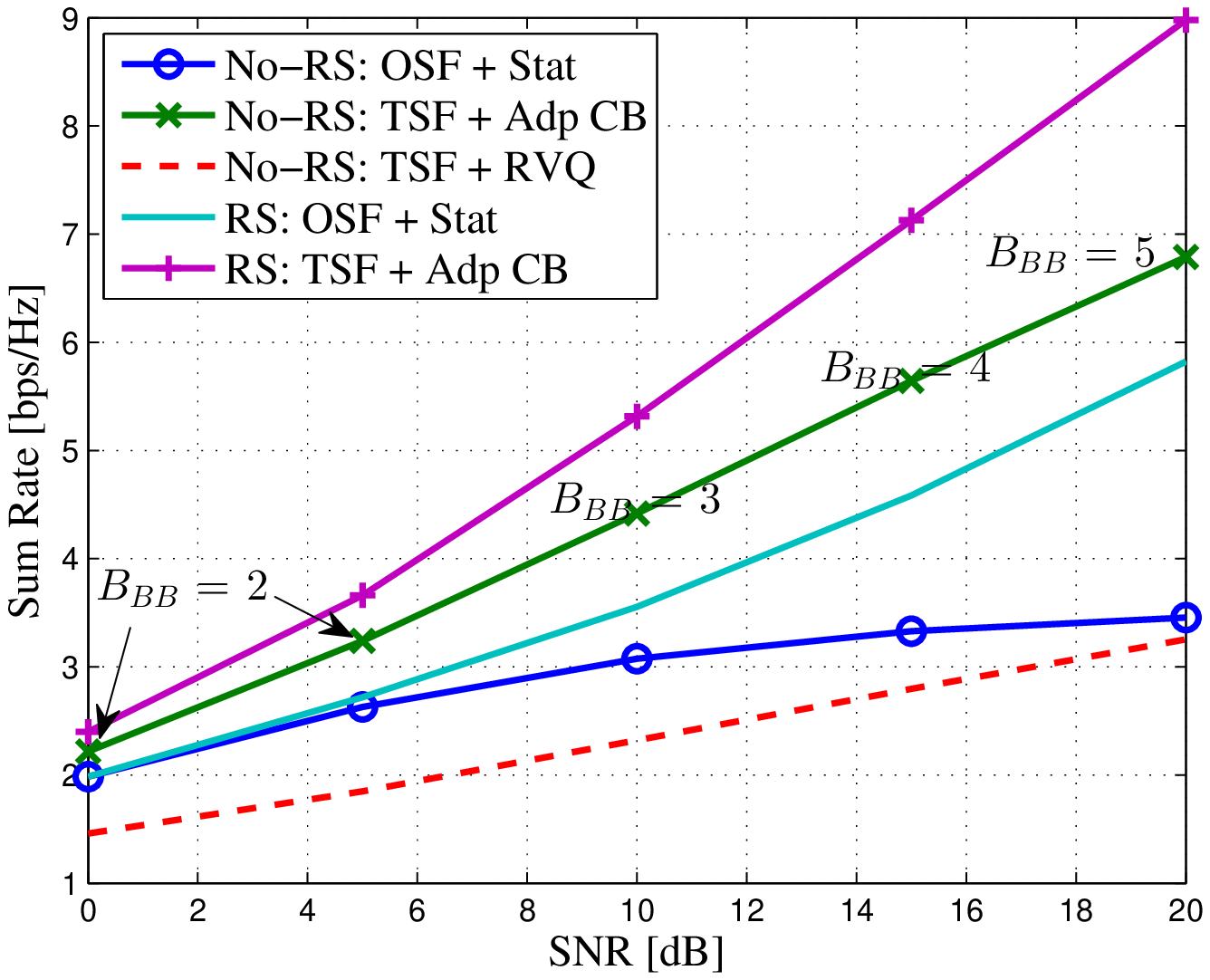}  } \hspace{-5pt}
\subfigure[]{ \label{fig8} \includegraphics[width = 0.45\textwidth]{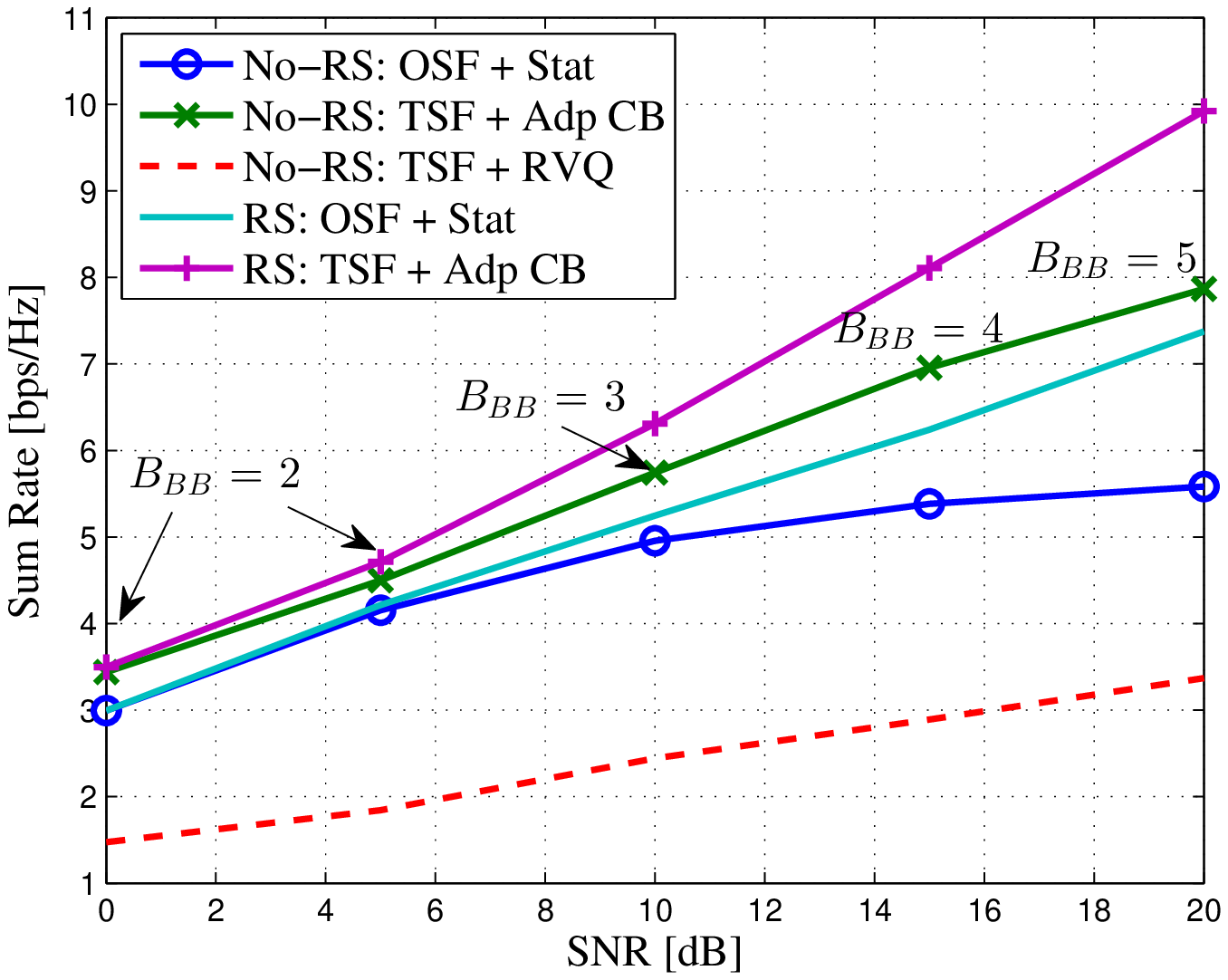} }
\caption{Comparison between No-RS with extra feedback and RS, $M = 32, K = 4, B_{RF} = 4$; (a) SBF/ZF digital precoder, (b) SLNR-based digital precoder.} \label{fig78}
\end{figure*}

In Fig. \ref{fig4}, the `OSF + Stat + SLNR' scheme slightly exceeds various baselines even in not-very-sparse ($L \gg 1$) channels, which confirms the effectiveness of the proposed design.
However, compared with Fig. \ref{fig3} where the channels are sparse ($L = 2$), the sum rates of the hybrid precoding schemes with $L = 15$ are highly degraded since one RF chain per user (i.e., $N = K$) captures only one path gain while missing the rest.
The sum rates are dominated by multiuser interference at high SNR since the covariance matrices of the effective channels with $L \gg 1$ tend to be full rank.

In Fig. \ref{fig5}, we evaluate the effect of the number of transmit antennas on the rate performance of the proposed hybrid precoding schemes at SNR $= -5$ dB.
The SLNR-based digital precoding is considered.
Each user channel has $L = 4$ paths and the total feedback overhead is $B = 8$.
A fixed feedback allocation is used for the `TSF + Adp CB' scheme, i.e., $B_{RF} = 5, B_{BB} = 3$.
Fig. \ref{fig5} shows that the rate gap between the `OSF + Stat' and `TSF + Adp CB' schemes enlarges as a larger number of antennas is employed at the BS, reaching as large as 3.9 bps/Hz at $M = 128$, SNR = -5 dB.
It indicates that the proposed `OSF + Stat' scheme works well in the large-scale array regime.
According to Section \ref{comp}, the `OSF + Stat' scheme forms a precise RF beamsteering with the help of all feedback resources.
Meanwhile, as $M$ becomes larger but $L = 4$ is fixed, the channel covariance matrices of users have higher dimensional nullspace to efficiently eliminate the multiuser interference.
On the contrary, the rate gain of the `TSF + Adp CB' scheme from increasing $M$ is small due to coarse (low resolution) analog beamforming.

In Fig. \ref{fig6}, we set $M = 32, L = 4$, SNR $= 5$ dB and evaluate the performance of the proposed schemes for varying $B$.
Fig. \ref{fig6} shows that when the amount of feedback overhead is small, it is preferable to allocate all resources to the analog beamforming (i.e., `OSF + Stat').
As $B$ increases, the feedback resources should be divided such that RF codebook with $B_{RF}$ has sufficient resolution to distinguish different channel paths.
Moreover, the remaining resources can accurately quantize the effective channel.
The digital precoder based on the accurate channel quantization effectively mitigates the multiuser interference.
In the large regime of feedback overhead $B$, Fig. \ref{fig6} shows that the `TSF + Adp CB' scheme is preferable over the `OSF + Stat' scheme, which validates the last observation in Section \ref{comp}.

\subsection{Rate Splitting with Hybrid Precoding}
%%%%%%%%%%%%%%%%%%%%%%%%%%%%%%%%%%%%%%%%%%%%%%%%%%%%%%%%%%%%%%%%%%%%%%%%%%%%%%%%%%%%%%%%%%%%%%%%%%%%%%%%%%%
In Fig. \ref{fig78}, we compare the `OSF + Stat' precoded RS with the `TSF + Adp CB' precoded No-RS.
Both strategies use the same $B_{RF}$ for the first-stage analog beamforming while `TSF + Adp CB' is driven by additional feedback $B_{BB}$ in the second-stage channel quantization.
We also plot the `OSF + Stat'/`TSF + RVQ' precoded No-RS and the `TSF + Adp CB' precoded RS as references.
The power splitting ratio in the `TSF + Adp CB' precoded RS is obtained by exhaustive search.
SBF/ZF and SLNR-based digital precoders are respectively considered in Fig. \ref{fig7} and Fig. \ref{fig8} for the proposed strategies.
The number of feedback bits in the first-stage beam steering is set as $B_{RF} = 4$ while $B_{BB}$ is annotated in Fig. \ref{fig78}.
Following Remark 2, the amount of extra feedback bits should scale as $B_{BB} = O(\frac{K - 1}{K}\frac{P_{dB}}{3})$ to achieve sum DoF of 1.

Fig. \ref{fig78} shows that the `OSF + Stat' precoded No-RS is interference-limited beyond SNR = 15 dB, since statistical CSIT-based digital precoding is unable to eliminate multiuser interference.
Meanwhile, the `TSF + Adp CB' or `TSF + RVQ' precoded No-RS strategies make use of extra SNR-adaptive second-stage feedback so as to achieve a rate performance that linearly increases with the transmit power.
We note that `OSF + Stat' precoded RS and `TSF + Adp CB'/`TSF + RVQ' precoded No-RS achieve almost the same sum DoF (reflected by the slope of the sum rate at high SNR).
Even though `TSF + Adp CB' precoded No-RS has a rate gain (around 1 bps/Hz) over `OSF + Stat' precoded RS, the latter enables saving both in the second-stage channel training and feedback.
Moreover, it can be seen that the sum rate of `OSF + Stat' precoded RS outperforms `TSF + RVQ' precoded No-RS \cite{Alk2015} due to the exploitation of the second-order statistics and the common message transmission.
Finally, implementing RS with `TSF + Adp CB' highly improves the achievable rate compared to the No-RS strategy.

%%%%%%%%%%%%%%%%%%%%%%%%%%%%%%%%%%%%%%%%%%%%%%%%%%%%%%%%%%%%%%%%%%%%%%%%%%%%%%%%%%%%%%%%%%%%%%%%%%%%%%%%%%%

\section{CONCLUSION} \label{conclusion}

In multiuser mm-Wave systems, a hybrid analog beamforming and digital precoding architecture is used to tackle the RF hardware constraint.
Considering the conventional multiuser transmission strategy, we proposed an `OSF + Stat' hybrid precoding scheme based on one-stage feedback and statistical CSIT.
Specifically, the `OSF + Stat' scheme uses all feedback resources to precisely design the first-stage analog beamformer while mitigating multiuser interference by statistical CSIT-based digital precoder.
As a comparison, the `TSF + Adp CB' scheme based on two-stage feedback partitions the total overhead into RF feedback and effective channel quantization.
The digital precoder copes with the multiuser interference based on quantized CSIT.
Under a fixed total feedback constraint, we showed that the `OSF + Stat' scheme outperforms the `TSF + Adp CB' scheme for very limited feedback and/or very sparse channels.
Meanwhile, the `OSF + Stat' scheme gets rid of the second-stage channel training and feedback.

Nevertheless, we note that the conventional transmission strategy precoded by either `OSF + Stat' or `TSF + Adp CB' is interference-limited at high SNR.
Then, we proposed a rate splitting transmission strategy which tackles the residual interference.
The idea of RS can be applied with both one-stage/two-stage hybrid precoding schemes to enhance the rate performance.
In consideration of the benefits of RS, we particularly showed that given the same amount of feedback for the first-stage analog beamforming, the `OSF + Stat' precoded RS can achieve a rate performance comparable to that of the `TSF + Adp CB' precoded No-RS with extra second-stage feedback for channel quantization.
By employing a more sophisticated transceiver architecture (i.e., superposition coding at the transmitter and SIC at the receiver), the RS transmission strategy enables significant saving in the second-stage channel training and feedback.

\appendix
%\section{PROOFS OF PROPOSITIONS AND COROLLARIES} \label{sec:Appendix}

\subsection{Proof of Proposition 1} \label{sec:prop1}
Let us first consider the non-overlapping case, i.e., $\|\mathbf{d}_k + \mathbf{d}_j\|_0 = 2L, \forall j \ne k$.
Given DFT codebook and $B = \log_2(M)$, the RF beamformer of user $k$ in \eqref{eq:beam} that maximizes the signal power is given by $\mathbf{f}_k = \mathbf{e}_{k,1}$.
Then, we have $\mathbf{F} = [\mathbf{e}_{1,1}, \cdots, \mathbf{e}_{K,1}]$ and $\mathbf{R}_{k,\text{eff}} = \mathbf{F}^H \mathbf{R}_{k} \mathbf{F} = \text{diag}\{\mathbf{1}_k\}$.
A straightforward calculation of \eqref{eq:GE} gives $\mathbf{w}_k = \mathbf{w}^{\star}_k = \mathbf{1}_k$.
Noting that $\mathbf{h}_k = \sqrt{\frac{M}{L}} \sum^{L}_{l=1} g_{k,l} \, \mathbf{e}_{k,l}$, the achievable rate of user $k$ is thus given by
\begin{eqnarray} \label{eq:rateapp1}
R_k &=& \log_2 \Big(1 + \frac{\rho |\mathbf{h}^H_{k} \mathbf{f}_{k} |^2 }{1+ \rho \sum_{j \ne k} |\mathbf{h}^H_{k} \mathbf{f}_{j} |^2}\Big)  \\
&=& \log_2 \Big(1 +\rho \frac{M}{L} |g_{k,1}|^2 \Big),
\end{eqnarray}
which is obtained by the fact that $\mathbf{f}_j = \mathbf{e}_{j,1} \perp \mathbf{e}_{k,l}, \forall k \ne j$ in the non-overlapped scenario.

Secondly, we consider the fully overlapped case where all users share the identical steering vectors set $\{\mathbf{e}_{l}\}$ and therefore the same channel covariance matrix $ \mathbf{R}_k = \mathbf{R} =  \frac{M}{L} \sum^L_{l=1} \mathbf{e}_l \mathbf{e}_l^H$.
Without loss of generality, we have $\mathbf{f}_k \in \mathbf{e}_{l = 1,\cdots, L}$ with $\mathbf{f}_k \ne \mathbf{f}_{j|j \ne k}$ from \eqref{eq:beam} and $\mathbf{R}_{k,\text{eff}} = \mathbf{F}^H \mathbf{R}_{k} \mathbf{F} = \frac{M}{L} \mathbf{I}_K$.
It can also be obtained that $\mathbf{w}_k = \mathbf{w}^{\star}_k = \mathbf{1}_k$.
Noting that $\mathbf{h}_k = \sqrt{\frac{M}{L}} \sum^{L}_{l=1} g_{k,l} \, \mathbf{e}_{k,l}$ with $\forall \mathbf{e}_{k,l} \in \mathbf{e}_{l}$, the achievable rate of user $k$ is given by
\begin{eqnarray}
R_k &=& \log_2 \Big(1 + \frac{\rho |\mathbf{h}^H_{k} \mathbf{f}_{k} |^2 }{1+ \rho \sum_{j \ne k} |\mathbf{h}^H_{k} \mathbf{f}_{j} |^2}\Big) \\
&=& \log_2 \Big(1 + \frac{\rho\frac{M}{L} |g_{k,1}|^2 }{1 + \rho\frac{M}{L} \sum_{j \ne k} |g_{k,l_j(l_j \ne 1)}|^2} \Big), \label{eq:rateapp2}
\end{eqnarray}
where $\mathbf{f}_j$ corresponds to $\mathbf{e}_{k,l_j}$.
Then, \eqref{eq:rateapp2} can be further lower bounded by \eqref{eq:achrate2} and the equality follows $L = K$.

\subsection{Proof of Proposition 3} \label{sec:prop3}
The average sum rate of RS can be written as
\begin{eqnarray} \label{eq:prop3-1}
\mathbb{E} (R^{RS}_{\text{sum}}) = \mathbb{E} (R^{c}) + \sum^K_{k=1} \mathbb{E} (R^p_k).
\end{eqnarray}

We first compute the average rate of the private message intended to user $k$ as
\begin{eqnarray} \label{eq:prop3-2}
\mathbb{E} (R^p_k) \!\!\!\!\!&=&\!\!\!\!\! \mathbb{E} \Big[ \log_2 \Big(1 +  \frac{P_{k} \, |\mathbf{h}^H_{k} \mathbf{F} \mathbf{w}_{k} |^2 }{1 + \sum_{j \ne k} P_{j} \, |\mathbf{h}^H_{k} \mathbf{F} \mathbf{w}_{j} |^2} \Big)\Big] \nonumber \\
&=& \!\!\!\!\! \mathbb{E} \Big[ \log_2 \Big(1 +  \exp \Big( \ln \frac{P_{k} \, |\mathbf{h}^H_{k} \mathbf{F} \mathbf{w}_{k} |^2 }{1 + \sum_{j \ne k} P_{j} \, |\mathbf{h}^H_{k} \mathbf{F} \mathbf{w}_{j} |^2} \Big) \Big)\Big] \nonumber \\
&\ge& \!\!\!\!\!  \log_2 \Big[1 +  \exp \Big( \mathbb{E} \ln (P_{k} \, |\mathbf{h}^H_{k} \mathbf{F} \mathbf{w}_{k} |^2)  - \nonumber \\
&& \qquad \qquad \quad \;\, \mathbb{E} \ln (1 + \sum_{j \ne k} P_{j} \, |\mathbf{h}^H_{k} \mathbf{F} \mathbf{w}_{j} |^2) \Big) \Big] \label{prop3-21} \\
&\ge& \!\!\!\!\!  \log_2 \Big[1 +  \exp \Big( \mathbb{E} \ln (P_{k} \, |\mathbf{h}^H_{k} \mathbf{F} \mathbf{w}_{k} |^2) \nonumber \\
&& \qquad \;  - \ln (1 + \sum_{j \ne k} P_{j} \,\mathbf{w}^H_j \mathbf{R}_{k,\text{eff}} \mathbf{w}_j ) \Big) \Big] \label{prop3-22} \\
&=& \!\!\!\!\!  \log_2 \Big[1 +  \exp \Big( \ln (P_{k} \,\mathbf{w}^H_k \mathbf{R}_{k,\text{eff}} \mathbf{w}_k) - \gamma \nonumber \\
&& \qquad \;  - \ln (1 + \sum_{j \ne k} P_{j} \,\mathbf{w}^H_j \mathbf{R}_{k,\text{eff}} \mathbf{w}_j ) \Big) \Big]  \label{prop3-23} \\
&=& \log_2 \bigg(1 + \frac{e^{-\gamma} \cdot \frac{Pt}{K} \mathbf{w}^H_k \mathbf{R}_{k,\text{eff}} \mathbf{w}_k }{1 + \frac{Pt}{K} \sum_{j \ne k} \mathbf{w}^H_j \mathbf{R}_{k,\text{eff}} \mathbf{w}_j} \bigg), \label{prop3-24}
\end{eqnarray}
where \eqref{prop3-21} is due to the convexity of $\log_2(1 + e^x)$ in $x$ while \eqref{prop3-22} is obtained by applying Jensen's inequality.
Define the rank one matrix $\mathbf{X}_k \triangleq \mathbf{A}_k^H \mathbf{F} \mathbf{w}_{k} \mathbf{w}^H_{k} \mathbf{F}^H \mathbf{A}_k$ and decompose it as $\mathbf{X}_k = \mathbf{U}_k \mathbf{\Lambda}_k \mathbf{U}^H_k$.
Then, we have $|\mathbf{h}^H_{k} \mathbf{F} \mathbf{w}_{k} |^2 =  \mathbf{g}^H_{k} \mathbf{X}_k  \mathbf{g}_{k} \overset{d}{=} \mathbf{g}^H_{k} \mathbf{\Lambda}_k  \mathbf{g}_{k} = \lambda_k |g_{k,m}|^2$, where $\overset{d}{=}$ indicates the equivalence in distribution and $\lambda_k$ is the only non-zero entry of $\mathbf{\Lambda}_k$.
Since $|g_{k,m}|^2 \sim \text{Exp}(1)$ and $\lambda_k = \text{tr}(\mathbf{\Lambda}_k ) = \text{tr}(\mathbf{X}_k) = \mathbf{w}^H_k \mathbf{R}_{k,\text{eff}} \mathbf{w}_k$, we can obtain \eqref{prop3-23} by $ \mathbb{E} [\ln (\mathbf{g}^H_{k} \mathbf{X}_k  \mathbf{g}_{k})] = \ln(\mathbf{w}^H_k \mathbf{R}_{k,\text{eff}} \mathbf{w}_k) - \gamma$, where $\gamma$ is the Euler constant.

A direct calculation of $\mathbb{E} (R^{c}) = \mathbb{E} [\underset{k}{\min} \; (R^c_k)]$ is technically challenging due to the requirement of the distributions of $R^c_k$ and further $\underset{k}{\min} \; (R^c_k)$.
We assume that $\mathbb{E} [\underset{k}{\min} \; (R^c_k)]$ can be well approximated by $\underset{k}{\min} \; \mathbb{E} (R^c_k)$.
By following a similar derivation, we can compute the average rate of the common message seen by user $k$ as
\begin{eqnarray} \label{eq:prop3-3}
\mathbb{E} (R^c_k) \!\!&=&\!\! \mathbb{E} \Big[ \log_2 \Big(1 +  \frac{P_{c} \, |\mathbf{h}^H_{k} \mathbf{F} \mathbf{w}_{c} |^2 }{1 + \sum^K_{j =1 } P_{j} \, |\mathbf{h}^H_{k} \mathbf{F} \mathbf{w}_{j} |^2} \Big)\Big] \\
&\ge& \!\!\!\! \log_2 \bigg(1 + \frac{e^{-\gamma}\cdot P(1-t)\, \mathbf{w}^H_c \mathbf{R}_{k,\text{eff}} \mathbf{w}_c}{1 +\frac{Pt}{K} \sum^K_{j=1} \mathbf{w}^H_j \mathbf{R}_{k,\text{eff}} \mathbf{w}_j} \bigg).
\end{eqnarray}

Finally, combining \eqref{prop3-24} and \eqref{eq:prop3-3} completes the proof.
Moreover, the effectiveness of Proposition 3 is also supported by \cite[Lemma 1]{zhang2014}.
It states that if $X = \sum_{i=1}^{n_1} x_i, Y = \sum_{j=1}^{n_2} y_j$ with random variables $x_i, y_j \in \mathbb{R}_{\ge0}$, we get $\mathbb{E}[\log_2 (1 + X/Y)] \approx \log_2 [1 + \mathbb{E}(X)/ \mathbb{E}(Y)]$ and the approximation error decreases as the number of random variables $n_1$ and $n_2$ increases.
It provides a useful reference calculation of the average rate which can be well approximated by $\mathbb{E}[\log_2 (1 + S/I)] \approx \log_2 [1 + \mathbb{E}(S)/ \mathbb{E}(I)]$, where $S$ and $I$ represent the signal power and interference plus noise, respectively.
Indeed, this average rate approximation is lower bounded by $\log_2 [1 + e^{-\gamma} \mathbb{E}(S)/ \mathbb{E}(I)]$ derived in Proposition 3, where $e^{-\gamma} \approx 0.56$.

\bibliographystyle{IEEEtran}
\bibliography{reference}

\begin{thebibliography}{10}
\providecommand{\url}[1]{#1}
\csname url@samestyle\endcsname
\providecommand{\newblock}{\relax}
\providecommand{\bibinfo}[2]{#2}
\providecommand{\BIBentrySTDinterwordspacing}{\spaceskip=0pt\relax}
\providecommand{\BIBentryALTinterwordstretchfactor}{4}
\providecommand{\BIBentryALTinterwordspacing}{\spaceskip=\fontdimen2\font plus
\BIBentryALTinterwordstretchfactor\fontdimen3\font minus
  \fontdimen4\font\relax}
\providecommand{\BIBforeignlanguage}[2]{{%
\expandafter\ifx\csname l@#1\endcsname\relax
\typeout{** WARNING: IEEEtran.bst: No hyphenation pattern has been}%
\typeout{** loaded for the language `#1'. Using the pattern for}%
\typeout{** the default language instead.}%
\else
\language=\csname l@#1\endcsname
\fi
#2}}
\providecommand{\BIBdecl}{\relax}
\BIBdecl

\bibitem{Rappaport2013}
T.~S. Rappaport, S.~Sun, R.~Mayzus, H.~Zhao, Y.~Azar, K.~Wang, G.~N. Wong,
  J.~K. Schulz, M.~Samimi, and F.~Gutierrez, ``Millimeter wave mobile
  communications for {5G} cellular: It will work!'' \emph{IEEE Access}, vol.~1,
  pp. 335--349, 2013.

\bibitem{andrews2014}
J.~G. Andrews, S.~Buzzi, W.~Choi, S.~Hanly, A.~Lozano, A.~C. Soong, and J.~C.
  Zhang, ``What will {5G} be?'' \emph{IEEE J. Select. Areas Commun.}, vol.~32,
  no.~6, June 2014.

\bibitem{Pi2011}
Z.~Pi and F.~Khan, ``An introduction to millimeter-wave mobile broadband
  systems,'' \emph{IEEE Commun. Mag.}, vol.~49, no.~6, pp. 101--107, 2011.

\bibitem{Omar2014}
O.~E. Ayach, S.~Rajagopal, S.~Abu-Surra, Z.~Pi, and R.~W. Heath, ``Spatially
  sparse precoding in millimeter wave {MIMO} systems,'' \emph{IEEE Trans.
  Wireless Commun.}, vol.~13, no.~3, pp. 1499--1513, March 2014.

\bibitem{Gaoz2015}
Z.~Gao, L.~Dai, D.~Mi, Z.~Wang, M.~A. Imran, and M.~Z. Shakir, ``Mmwave
  massive-{MIMO}-based wireless backhaul for the {5G} ultra-dense network,''
  \emph{IEEE Wireless Commun.}, vol.~22, no.~5, pp. 13--21, Oct. 2015.

\bibitem{Yu2016}
X.~Yu, J.~C. Shen, J.~Zhang, and K.~Letaief, ``Alternating minimization
  algorithms for hybrid precoding in millimeter wave {MIMO} systems,''
  \emph{IEEE J. Sel. Topics Signal Process.}, vol.~10, no.~3, pp. 485--500,
  April 2016.

\bibitem{Foad2016}
F.~Sohrabi and W.~Yu, ``Hybrid digital and analog beamforming design for
  large-scale antenna arrays,'' \emph{IEEE J. Sel. Topics Signal Process.},
  vol.~10, no.~3, pp. 501--513, April 2016.

\bibitem{Liang2014}
L.~Liang, W.~Xu, and X.~Dong, ``Low-complexity hybrid precoding in massive
  multiuser {MIMO} systems,'' \emph{IEEE Wireless Commun. Lett.}, vol.~3,
  no.~6, pp. 653--656, 2014.

\bibitem{Gao2015}
X.~Gao, L.~Dai, S.~Han, C.-L. I, and R.~W. Heath, ``Energy-efficient hybrid
  analog and digital precoding for mmwave {MIMO} systems with large antenna
  arrays,'' \emph{IEEE J. Sel. Areas Commun.}, vol.~34, no.~4, pp. 998--1009,
  April 2016.

\bibitem{Alk2014}
A.~Alkhateeb, O.~E. Ayach, G.~Leus, and R.~W. Heath, ``Channel estimation and
  hybrid precoding for millimeter wave cellular systems,'' \emph{IEEE J. Sel.
  Topics Signal Process.}, vol.~8, no.~5, pp. 831--846, 2014.

\bibitem{Bog2015}
T.~E. Bogale, L.~B. Le, and X.~Wang, ``Hybrid analog-digital channel estimation
  and beamforming: Training-throughput tradeoff,'' \emph{IEEE Trans. Commun.},
  vol.~63, no.~12, pp. 5235--5249, 2015.

\bibitem{Alk2015}
A.~Alkhateeb, G.~Leus, and R.~W. Heath, ``Limited feedback hybrid precoding for
  multi-user millimeter wave systems,'' \emph{IEEE Trans. Wireless Commun.},
  vol.~14, no.~11, pp. 6481--6494, 2015.

\bibitem{Hao2013}
C.~Hao and B.~Clerckx, ``{MISO} broadcast channel with imperfect and
  (un)matched {CSIT} in the frequency domain: {DoF} region and transmission
  strategies,'' in \emph{Proc. IEEE PIMRC}, Sept. 2013, pp. 1--6.

\bibitem{Clerckx2016}
B.~Clerckx, H.~Joudeh, C.~Hao, M.~Dai, and B.~Rassouli, ``Rate splitting for
  {MIMO} wireless networks: a promising {PHY}-layer strategy for {LTE}
  evolution,'' \emph{IEEE Commun. Mag.}, vol.~54, no.~5, pp. 98--105, May 2016.

\bibitem{Dai2016}
M.~Dai, B.~Clerckx, D.~Gesbert, and G.~Caire, ``A rate splitting strategy for
  massive {MIMO} with imperfect {CSIT},'' \emph{IEEE Trans. Wireless Commun.},
  vol.~15, no.~7, pp. 4611--4624, July 2016.

\bibitem{Hao2016-1}
C.~Hao and B.~Clerckx, ``{MISO} networks with imperfect {CSIT}: A topological
  rate-splitting approach,'' \emph{IEEE Trans. Commun.}, vol.~PP, no.~99, pp.
  1--1, 2017.

\bibitem{Hao2016-2}
\BIBentryALTinterwordspacing
C.~Hao, B.~Rassouli, and B.~Clerckx, ``Achievable {DoF} regions of {MIMO}
  networks with imperfect {CSIT},'' \emph{submitted to IEEE Trans. Inf.
  Theory}, March 2016. [Online]. Available:
  \url{https://arxiv.org/pdf/1603.07513v2.pdf}
\BIBentrySTDinterwordspacing

\bibitem{Hamdi2016-3}
H.~Joudeh and B.~Clerckx, ``A rate-splitting strategy for max-min fair
  multigroup multicasting,'' in \emph{Proc. IEEE 17th Int. Workshop Signal
  Process. Adv. Wireless Commun. (SPAWC)}, July 2016, pp. 1--5.

\bibitem{Hamdi2016-1}
------, ``Sum-rate maximization for linearly precoded downlink multiuser {MISO}
  systems with partial csit: A rate-splitting approach,'' \emph{IEEE Trans.
  Commun.}, vol.~64, no.~11, pp. 4847--4861, Nov 2016.

\bibitem{Hamdi2016-2}
------, ``Robust transmission in downlink multiuser {MISO} systems: A
  rate-splitting approach,'' \emph{IEEE Trans. Signal Process.}, vol.~64,
  no.~23, pp. 6227--6242, Dec. 2016.

\bibitem{Say2007}
A.~M. Sayeed and V.~Raghavan, ``Maximizing {MIMO} capacity in sparse multipath
  with reconfigurable antenna arrays,'' \emph{IEEE J. Sel. Topics Signal
  Process.}, vol.~1, no.~1, pp. 156--166, June 2007.

\bibitem{Mo2014}
J.~Mo and R.~W. Heath, ``High {SNR} capacity of millimeter wave {MIMO} systems
  with one-bit quantization,'' in \emph{Proc. IEEE Information Theory and
  Applications Workshop (ITA)}, Feb. 2014, pp. 1--5.

\bibitem{Adh2014}
A.~Adhikary, E.~A. Safadi, M.~K. Samimi, R.~Wang, G.~Caire, T.~S. Rappaport,
  and A.~F. Molisch, ``Joint spatial division and multiplexing for mm-wave
  channels,'' \emph{IEEE J. Sel. Areas Commun.}, vol.~32, no.~6, pp.
  1239--1255, June 2014.

\bibitem{Xiao2016}
Z.~Xiao, T.~He, P.~Xia, and X.~G. Xia, ``Hierarchical codebook design for
  beamforming training in millimeter-wave communication,'' \emph{IEEE Trans.
  Wireless Commun.}, vol.~15, no.~5, pp. 3380--3392, May 2016.

\bibitem{Sadek2007}
M.~Sadek, A.~Tarighat, A.~H. Sayed \emph{et~al.}, ``A leakage-based precoding
  scheme for downlink multi-user {MIMO} channels,'' \emph{IEEE Trans. Wireless
  Commun.}, vol.~6, no.~5, pp. 1711--1721, May 2007.

\bibitem{Raghavan2011}
V.~Raghavan, S.~Hanly, and V.~Veeravalli, ``Statistical beamforming on the
  {G}rassmann manifold for the two-user broadcast channel,'' \emph{IEEE Trans.
  Inf. Theory}, vol.~59, no.~10, pp. 6464--6489, Oct. 2013.

\bibitem{Dai2015}
M.~Dai and B.~Clerckx, ``Transmit beamforming for {MISO} broadcast channels
  with statistical and delayed {CSIT},'' \emph{IEEE Trans. Commun.}, vol.~63,
  no.~4, pp. 1202--1215, April 2015.

\bibitem{Bruno2013}
B.~Clerckx and C.~Oestges, \emph{{MIMO} Wireless Networks: Channels, Techniques
  and Standards for Multi-antenna, Multi-user and Multi-cell Systems}.\hskip
  1em plus 0.5em minus 0.4em\relax Academic Press (Elsevier), 2013.

\bibitem{Love2004}
D.~J. Love, R.~W. Heath, W.~Santipach, and M.~L. Honig, ``What is the value of
  limited feedback for {MIMO} channels?'' \emph{IEEE Commun. Mag.}, vol.~42,
  no.~10, pp. 54--59, 2004.

\bibitem{Bruno2008}
B.~Clerckx, G.~Kim, and S.~Kim, ``{MU-MIMO} with channel statistics-based
  codebooks in spatially correlated channels,'' in \emph{Proc. IEEE GLOBECOM},
  2008, pp. 1--5.

\bibitem{Choi2013}
J.~Choi, V.~Raghavan, and D.~J. Love, ``Limited feedback design for the
  spatially correlated multi-antenna broadcast channel,'' in \emph{Proc. IEEE
  GLOBECOM}, 2013, pp. 3481--3486.

\bibitem{Love2006}
D.~J. Love and R.~W. Heath, ``Limited feedback diversity techniques for
  correlated channels,'' \emph{IEEE Trans. Veh. Technol.}, vol.~55, no.~2, pp.
  718--722, March 2006.

\bibitem{Shen2016}
W.~Shen, L.~Dai, Y.~Zhang, J.~Li, and Z.~Wang, ``On the performance of channel
  statistics-based codebook for massive {MIMO} channel feedback,'' \emph{IEEE
  Trans. Veh. Technol.}, vol.~PP, no.~99, pp. 1--1, 2017.

\bibitem{Jindal2006}
N.~Jindal, ``{MIMO} broadcast channels with finite-rate feedback,'' \emph{IEEE
  Trans. Inform. Theory}, vol.~52, no.~11, pp. 5045--5060, Nov. 2006.

\bibitem{Bocc2012}
F.~Boccardi, B.~Clerckx, A.~Ghosh, E.~Hardouin, G.~Jongren, K.~Kusume,
  E.~Onggosanusi, and Y.~Tang, ``Multiple antenna techniques in
  {LTE}-advanced,'' \emph{IEEE Comm. Magazine}, vol.~50, no.~2, pp. 114--121,
  Mar. 2012.

\bibitem{Lim2013}
C.~Lim, T.~Yoo, B.~Clerckx, B.~Lee, and B.~Shim, ``Recent trend of multiuser
  {MIMO} in {LTE}-advanced,'' \emph{IEEE Comm. Magazine}, vol.~51, no.~3, pp.
  127--135, Mar. 2013.

\bibitem{Tom2006}
N.~Sidiropoulos, T.~Davidson, and Z.-Q. Luo, ``Transmit beamforming for
  physical-layer multicasting,'' \emph{IEEE Trans. Signal Process.}, vol.~54,
  no.~6, pp. 2239--2251, June 2006.

\bibitem{Meh2015}
O.~Mehanna, K.~Huang, B.~Gopalakrishnan, A.~Konar, and N.~D. Sidiropoulos,
  ``Feasible point pursuit and successive approximation of non-convex
  {QCQP}s,'' \emph{IEEE Signal Process. Lett.}, vol.~22, no.~7, pp. 804--808,
  July 2015.

\bibitem{Beck2010}
A.~Beck, A.~Ben-Tal, and L.~Tetruashvili, ``A sequential parametric convex
  approximation method with applications to nonconvex truss topology design
  problems,'' \emph{J. Global Optim.}, vol.~47, no.~1, pp. 29--51, 2010.

\bibitem{Hao2015}
C.~Hao, Y.~Wu, and B.~Clerckx, ``Rate analysis of two-receiver {MISO} broadcast
  channel with finite rate feedback: A rate-splitting approach,'' \emph{IEEE
  Trans. Commun.}, vol.~63, no.~9, pp. 3232--3246, Sept. 2015.

\bibitem{zhang2014}
Q.~Zhang, S.~Jin, K.-K. Wong, H.~Zhu, and M.~Matthaiou, ``Power scaling of
  uplink massive mimo systems with arbitrary-rank channel means,'' \emph{IEEE
  J. Sel. Topics Signal Process.}, vol.~8, no.~5, pp. 966--981, Oct. 2014.

\end{thebibliography}

\end{document}